\documentstyle[pre,eqsecnum,aps,epsf]{revtex}
\begin{document}
\draft
\title{Energy Transport in the Integrable System  
in Contact with Various Types of Phonon Reservoirs}
\author{K. Saito}
\address{
Department of Applied Physics, School of Engineering \\ 
University of Tokyo, Bunkyo-ku, Tokyo 113--8656,
Japan}
\author{S. Takesue}
\address{
Faculty of Integrated Human Studies,
Kyoto University, \\
Kyoto 606--8501, Japan}
\author{S. Miyashita}
\address{
Department of Applied Physics, School of Engineering \\ 
University of Tokyo, Bunkyo-ku, Tokyo 113--8656, Japan}
\date{\today}
\maketitle
\begin{abstract}
We study how energy transport in an integrable system is affected by the
spectral densities of heat reservoirs. The model investigated here is the
quantum harmonic chain whose both ends are in contact with two heat
reservoirs at different temperatures. The master equation for the reduced
density matrix is derived on the assumption that the reservoirs are composed
of an infinite number of independent harmonic oscillators.  We evaluate 
temperature profile and energy flux in the stationary state for the master
equation and discuss how they depend on the types of spectral densities.
When we attach the reservoirs of the same type of spectral density, 
we find that the temperature profile is independent of 
the types. On the other hand,
when the two reservoirs have different types of spectral densities, 
the energy profile near the ends of the chain depends on the types.
When the coupling is finite, the temperature profile 
near the ends shows wide variation of behavior dependent on 
spectral densities and temperatures of reservoirs.
This dependence is discussed with the Fokker-Planck equations
obtained in the classical limit.
\end{abstract}
\quad\\

\section{Introduction}
Generally integrable systems show abnormal energy transport, namely the
Fourier heat law is not realized there.  This attributes to the lack of
scattering between modes, which should be induced by nonintegrability.
Two typical characteristics are seen in the energy transport in integrable
systems. One is energy flux per unit volume independent of the system size. 
The other is a flat temperature profile with no global temperature gradient.

The harmonic chain is a typical integrable system which shows these 
characteristics \cite{RLL67,ZT90,IM70,V71,VR75}.
Rieder, Lebowitz and Lieb (RLL) investigated the classical
harmonic chain whose ends are in contact with heat reservoirs at different
temperatures \cite{RLL67}. 
They exactly evaluated the covariance matrix of the variables 
in the stationary state using the Langevin equation, and 
they proved these characteristics. That is, they found that energy flux 
per volume is proportional to only temperature difference and 
is independent of the system size, and no global temperature gradient is 
formed. Although the temperature profile is flat in the internal region,
they found the peculiar behavior in the vicinity of the ends of the chain.
Namely, the local temperature is {\em higher} than the bulk value near the
{\em colder} reservoir, and {\em lower} near the {\em hotter} reservoir.

Z\"{u}rcher and Talkner (ZT) \cite{ZT90} investigated a quantum model 
corresponding to that of RLL with use
of the quantum Langevin equation\cite{FKM65}.  As for the bulk behavior, 
they found the same features as in the classical case.
That is, 
no global temperature gradient is formed and energy flux is 
independent of system size. In the high temperature limit, the quantum 
Langevin equation is reduced to the Langevin equation with the 
Gaussian white noise and all the characteristics obtained in \cite{RLL67} 
are reproduced.
However, temperature profile in the vicinity of the
ends of the system shows some variety depending on temperature and a 
damping constant. 

The reservoir employed in these studies is only the Ohmic
type, which is one of the possible three types of heat reservoirs:
sub-Ohmic, Ohmic, and super-Ohmic. 
Because integrable systems have no scattering between modes, their 
nonequilibrium behavior will be easily affected by the types of reservoirs 
at the boundary. 
Thus in this paper, we investigate how nonequilibrium nature in 
the harmonic chain depends on the types of reservoirs.
Here we derive the master equation for
the reduced density matrix through the projection operator 
method on assumption that the reservoir is composed of 
an infinite number of independent harmonic oscillators.
This method is valid in the weak coupling limit and for slow 
motion of the system because it treats the second order perturbation with 
respect to a coupling constant and the Markovian approximation \cite{KTH85}.

We investigate the effect of the spectral density on
the temperature profile and energy flux 
in the quantum harmonic chain in contact with two reservoirs
applying the master equation for reduced density matrix.
At the weak coupling limit,
the stationary state can be obtained analytically, and it is found that
when the two reservoirs have the same type of spectral density, 
the temperature profile is independent of it, and the profile
is the same as the result previously reported\cite{RLL67,ZT90}.
On the other hand, when the spectral densities are of different types, 
the temperature profile depends on the types.
Even in the classical limit, 
the internal temperature deviates from the mean value 
of the temperature of the reservoirs and we observe a deviation of
temperatures around ends from the temperature of internal region.
The direction of deviation is 
determined by frequency dependence of the spectral density near $\omega = 0$

We also investigate the reduced density matrix with finite values of 
the coupling constant numerically. Although it is derived in the 
perturbation of the coupling constant and it is only valid in the weak 
coupling limit,
we dare regard the master equation with a finite coupling constant 
as a model for a time evolution with a dissipation. In other words, 
we assume that the time evolution qualitatively represents a kind of real
phenomena in nature, where the coupling constant represents the strength 
of the dissipative mechanism. We study qualitatively 
the dependence of temperature profile on the types of reservoirs.
It even has been reported in some cases that the reduced density 
matrix with a finite coupling can produce a good long time behavior 
of the system in comparison with the exact path-integral result 
not only qualitatively but also quantitatively \cite{KDH97}. 

Temperature profile at the ends of the
chain is found to depend on the spectral density and the
temperatures of the reservoirs even when reservoirs at both ends 
have same type of spectral density.
The dependence on the types of the reservoirs is discussed 
in the Fokker-Planck equations in the classical limit.
 
This paper is organized as follows. In Sec. II, we
derive the master equation for the reduced density matrix of a general 
many body system in contact with a phonon reservoir.
Sec. III is devoted to the investigation on energy transport
in the harmonic chain coupled to the phonon reservoirs at weak coupling limit.
In Sec. VI, we consider the case of finite coupling 
and investigate corresponding Fokker-Planck equations.
Summary and discussions are given in Sec. V.

\section{The Phonon Reservoir}
\subsection{Master equation for reduced density matrix}

In this subsection, we derive 
the master equation for the reduced density 
matrix of a system in contact with a phonon reservoir. 
Let us consider the following 
total Hamiltonian $H_{\rm tot}$,
\begin{equation}
H_{\rm tot} = H + H_{\rm Int} + H_{\rm R},
\end{equation}
where $H$ denotes the Hamiltonian for the system of interest, $H_{\rm R}$
denotes the Hamiltonian for the reservoir, and $H_{\rm Int}$ 
describes the interaction between the system and the reservoir.
We assume that the reservoir consists of an infinite number of mutually
independent harmonic oscillators \cite{CL83,GSI88,VS98}, that is,
\begin{equation}
H_{\rm R} = 
\sum_{\alpha } \frac{p_{\alpha}^{2}}{2 m_{\alpha}} 
+ \frac{m_{\alpha}\omega_{\alpha}^{2} x_{\alpha}^{2}}{2}  
= \sum_{\alpha} \hbar \omega_{\alpha}\left( b_{\alpha}^{\dagger}b_{\alpha} 
+\frac{1}{2}  \right), \label{RESER}
\end{equation}
where $b_{\alpha}^\dagger$ and $b_{\alpha}$ are the creation and annihilation
operators for the $\alpha$th mode. We assume a linear
coupling between a Hermitian operator of the system $X$ and 
reservoir's operators in the form.
\begin{eqnarray}
H_{\rm Int} &=& \sum_{\alpha }
\lambda 
\left( \frac{m_{\alpha}\omega_{\alpha}}{2}\right)^{1/2}
\gamma_{\alpha} x_{\alpha}X 
+ {\lambda '}^2 \sum_{\alpha } \frac{\gamma_{\alpha}^{2} }
                 { 4 \omega_{\alpha } } X^2  \nonumber \\
&=& \lambda 
\sqrt{\hbar}\sum_{\alpha} \gamma_\alpha 
\left(b_{\alpha}^{\dagger}+b_{\alpha}\right)X
+ {\lambda '}^2 \sum_{\alpha } \frac{\gamma_{\alpha}^{2} }
                 { 4 \omega_{\alpha } } X^2 \qquad ( \lambda ' \ge \lambda ),
\label{Hint}
\end{eqnarray}
where $\lambda $ is a coupling constant and $\gamma_\alpha$s and $\lambda '$
are some constants. We put the second term in the 
right hand side in order to make the total Hamiltonian to be bounded 
\cite{FLO88}. This term 
is regarded as a part of $H$:
\begin{eqnarray}
H & \rightarrow & H + {\lambda '}^2 \sum_{\alpha } \frac{\gamma_{\alpha}^{2} }
                 { 4 \omega_{\alpha } } X^2 . \label{RMH}
\end{eqnarray}
In this section, we do not make any assumption for the system, 
though we consider a harmonic chain for the system in the next section.

We derive the master equation 
for the reduced density matrix following the standard method \cite{KTH85}.
We start from the quantum Liouville equation for the total system
\begin{equation}
\frac{\partial\rho_{\rm tot}(t)}{\partial t}=
\frac{1}{i\hbar}[H_{\rm tot},\rho_{\rm tot}(t)], 
\end{equation}
where $\rho_{\rm tot}(t)$ is the density matrix for the total system.
Under the condition that the reservoir is initially in the equilibrium 
state at inverse temperature $\beta$, the degrees of freedom of
the reservoir are traced out with the aid of projection operators.  
In order to obtain an equation which can be solved practically,
we usually expand it up to the second order with respect to $\lambda$ 
and also adopt 
the Markovian approximation, which is valid when correlations between 
reservoir's variables are short-lived. 
As the result we obtain an equation for the reduced density
matrix $\rho(t)={\rm Tr_R}\rho_{\rm tot}(t)$ 
(${\rm Tr_R}$ means the trace concerning
the reservoir's degrees of freedom) of the form
\begin{equation}
\frac{\partial}{\partial t} \rho(t) = 
\frac{1}{i\hbar}\left[ H,\rho(t)\right] - \lambda^{2}\Gamma\rho(t),
\label{ME}
\end{equation}
where $\Gamma\rho(t)$ is given by
\begin{eqnarray}
\Gamma\rho(t) &=& \frac{1}{\hbar^{2}}
\int_{0}^{\infty}\, dt'\int_{-\infty }^{\infty }\, d\omega \,
e^{i\omega t'} \Phi(\omega)\left\{ XX(-t')\rho(t) \right. \nonumber \\
& & \left. \mbox{}- e^{\beta\hbar\omega}X\rho(t)X(-t') 
+ e^{\beta\hbar\omega}\rho(t)X(-t')X - X(-t')\rho(t)X
\right\} .
\label{DST}
\end{eqnarray}
In Eq.\ (\ref{DST}), $X(-t')$ means the Heisenberg operator at time $-t'$ 
\begin{equation}
X(-t') = e^{-iHt'/\hbar}X e^{iHt'/\hbar},
\end{equation}
and the function $\Phi(\omega )$ denotes the Fourier transform 
of the two-point function of the reservoir's 
operators coupling to $X$, namely,
\begin{equation}
\Phi(\omega)= 
\frac{1}{2\pi}\int_{-\infty}^{\infty}e^{-i\omega t}\Phi(t)dt,
\end{equation}
where $\Phi(t)$ is given by
\begin{eqnarray}
\Phi(t) &=& \left. {\rm Tr_R}\left[ 
\sum_{\alpha,\alpha'}
\left(\frac{m_{\alpha}\omega_{\alpha}}{2}\right)^{1/2}
\left(\frac{m_{\alpha'}\omega_{\alpha'}}{2}\right)^{1/2}
\gamma_\alpha\gamma_{\alpha'} x_{\alpha}(0)x_{\alpha'}(t)
e^{-\beta H_{\rm R}}\right]\right/{\rm Tr_R} e^{-\beta H_{\rm R}} \nonumber \\
&=& \hbar \sum_{\alpha }
\gamma_\alpha^2 \frac{e^{i \omega_{\alpha } t} 
+ e^{\beta \hbar \omega_{\alpha }} e^{- i \omega_{\alpha } t}  }
{e^{\beta \hbar \omega_{\alpha }} -1  }. \label{TPF}
\end{eqnarray}
Hence, denoting the reservoir's density of states with respect to 
frequency $\omega$ by $D(\omega)$, we can write 
$\Phi (\omega )$ as
\begin{eqnarray}
\Phi (\omega ) &=& \hbar\gamma(\omega)^2 \frac{D(\omega)-D(-\omega)}
{e^{\beta\hbar\omega }-1} ,
\label{phiomega}
\end{eqnarray}
where we introduced a smooth function $\gamma(\omega)$ that satisfies
$\gamma(\pm\omega_\alpha)=\gamma_\alpha$.
Here we define the spectral density $I(\omega )$ as
\begin{equation}
I(\omega ) = \gamma(\omega)^{2} D(\omega ) . 
\end{equation}
The following form for the spectral density is 
considered in the literature,
\begin{equation}
I(\omega) = I_{0}\omega^{\alpha} \theta (\omega ), \label{SPEDEF}
\end{equation} 
where $\theta (\omega )$ is the step function: 
$\theta (\omega ) = 1 \, {\rm for} \, \omega \ge 0$ and 
$\theta (\omega ) = 0 \, {\rm for} \, \omega < 0$.
The reservoir is called Ohmic if $\alpha = 1$,
sub-Ohmic if $\alpha<1$, and super-Ohmic if $\alpha > 1$\cite{GSI88}.

In the following we rewrite Eq.\ (\ref{ME}) in a form convenient
for later use. 
Let us consider the matrix components of operator $\Gamma\rho(t)$,
$\langle k |\Gamma\rho(t) | n \rangle $, 
where $|k \rangle$ and 
$| n \rangle $ are eigenstates for the system Hamiltonian $H$ with energy
eigenvalues $E_{k}$ and $E_{n}$, respectively.  
For the integral with 
respect to $t'$, we use the mathematical formula
\begin{equation}
\int_{0}^{\infty}e^{i\nu t}dt 
= \pi \delta(\nu ) + {\cal P}\frac{i}{\nu} ,
\end{equation}
neglecting the principal value \cite{KTH85,L73,WH65} and 
the Kubo-Martin-Schwinger (KMS) condition
$\Phi(\omega)e^{\beta\hbar\omega} =\Phi(-\omega)$. 
Then, the matrix components of operator $\Gamma\rho(t)$ is written as
\begin{eqnarray}
\langle k |\Gamma\rho(t)|n \rangle
&=&
\frac{\pi }{\hbar^{2}}\sum_{l,m}
\left[ X_{k,l}X_{l,m} 
\Phi\left( \frac{E_{l}-E_{m}}{\hbar} \right) \rho_{m,n} (t) 
 - X_{k,l}\rho_{l,m} (t) X_{n,m}^{*} 
\Phi \left( \frac{E_{n}-E_{m}}{\hbar}  \right) 
\right. \nonumber \\
&& \left. +\rho_{k,l} (t)X^{*}_{m,l}
\Phi \left( \frac{E_{m}-E_{l}}{\hbar} \right)X_{m,n} 
- X_{k,l}\Phi \left( \frac{E_{k}-E_{l}}{\hbar} \right) 
\rho_{l,m} (t)X_{m,n}
\right] .
\label{KN2}
\end{eqnarray}

Now we introduce the operator $R$ whose matrix elements are 
\begin{eqnarray}
\langle l | R | m \rangle &=& \frac{1}{\hbar} X_{l,m}
\Phi\left( \frac{E_{l}-E_{m}}{\hbar} \right) 
= 
\frac{1}{\hbar} X_{l,m} \frac{
  I( \frac{E_{l}-E_{m}}{\hbar} ) 
- I(- \frac{E_{l}-E_{m}}{\hbar} ) }
{e^{ \beta\left( E_{l}-E_{m} \right) } -1  } . 
\label{RFORM}
\end{eqnarray}
Then $\Gamma\rho(t)$ is written in the following compact form,
\begin{eqnarray}
\Gamma\rho(t) &=& \frac{\pi}{\hbar}
\left( XR\rho(t)-R\rho(t)X-X\rho(t)R^{\dagger}+\rho(t)R^{\dagger}X \right) 
\nonumber \\
&=& \frac{\pi}{\hbar}
\left(\left[X,R\rho(t)\right]+\left[X,R\rho(t)\right]^{\dagger}\right) ,
\end{eqnarray}
Thus we arrive at the master equation of the form
\begin{equation}
\frac{\partial\rho(t)}{\partial t}= 
\frac{1}{i\hbar}\left[H,\rho(t)\right] 
-\frac{\pi\lambda^2}{\hbar}
\left( \left[X,R\rho(t)\right] + \left[X,R\rho(t)\right]^{\dagger} \right). 
\label{CTTR}
\end{equation}
This is a generalized Lindblad form \cite{L76,KG97} 
treating general many body system with the coupling form (\ref{Hint}).
When the system has many body interaction, the noncommutabilities cause 
the operator $R$ to contain all degrees of freedom of the system even 
if $H_{\rm Int}$ is a part of the system. Thus, in general $R$ has a 
complicated form with all degrees freedom of the system. 
Nevertheless, Eq.\ (\ref{RFORM}) gives the explicit and compact form of $R$ 
for the general systems when the reservoir is given by (\ref{RESER}).
Thus we can expect that the master equation (\ref{CTTR}) is widely 
applicable for the concrete studies of many body systems.

In the present study this master equation is used as a basic equation 
for a system coupled with the phonon reservoir. It is readily checked that
Eq.\  (\ref{CTTR}) satisfies at least a necessary condition 
for the master equation, i.e., the canonical distribution
$e^{-\beta H}/{\rm Tr}(e^{-\beta H}) $ into $\rho$
in Eq.\ (\ref{CTTR}) gives a stationary solution.
We also expect the stability of the stationary solution 
at least when $\lambda$ is small enough.

\subsection{Comparison with the Quantum Langevin dynamics}

Here we briefly review another type of equation representing quantum dynamics 
with a thermal environment that is called a quantum Langevin equation
which was used in previous studies \cite{ZT90} 
and compare it with the master equation for reduced density 
matrix (\ref{CTTR}) (see also \cite{FLO88} for
other types of quantum Langevin equations we no not explain here ).

The quantum Langevin equation was introduced by Ford, Kac, and 
Mazur \cite{FKM65}. They considered the following coupled oscillators 
composed of $(2N+1)$ particles,
\begin{equation}
H = \frac{1}{2} \sum_{n=-N}^{N} p_{n}^{2} 
   + \frac{1}{2}\sum_{m,n=-N}^{N} q_{m}A_{m,n}q_{n},
\label{HFKM}
\end{equation}
where $q_{n}$ and $p_{n}$ are the $i$th canonical coordinate and momentum
variable, respectively. The matrix ${\bf A}=(A_{m,n})$ is a 
$(2N+1)\times(2N+1)$ symmetric matrix whose elements are 
\begin{eqnarray}
A_{m,n} &=& \frac{1}{2N+1}\sum_{k=-N}^{N}\omega_{k}^{2}
\exp \left[ i \frac{2\pi k}{2N+1}(m-n)   \right].
\end{eqnarray}
It should be noted that the eigenvalues of this matrix are
$\omega_{s}^{2}$ ($s =-N,-N+1,\cdots,N-1,N$).
The authors assumed that the initial state of the system $(\ref{HFKM})$
is in equilibrium at a temperature, and examined under what condition 
the behavior of particle $0$ can be described by a Langevin equation.

They found the following. If the eigen-frequencies of the 
whole system, $\omega_{s}$, have the special form
\begin{equation}
\omega_{s}^{2} = f^{2} \tan^{2} 
\left( \frac{\pi s}{2N+1} \right) ,
\end{equation}
the motion of a particle of the system in the equilibrium state 
is described by
\begin{mathletters}
\label{qleq}
\begin{eqnarray}
\frac{\partial q_{0}(t)}{\partial t} &=& p_{0} ,  \\
\frac{\partial p_{0}(t)}{\partial t} &=& - f p_{0} + E(t), 
\end{eqnarray}
\end{mathletters}
where $q_{0}, p_{0}$ and $E(t)$ are operators in the Heisenberg picture. 
The operator $E(t)$ is described by operators of particles.
In the equilibrium state, $E(t)$ behaves like the 
Gaussian random force with vanishing mean 
$\langle E(t)\rangle= 0$, where $\langle ... \rangle$ means ${\rm Tr}
\left( \exp (-\beta \, H( \{ q_{i} (0) \}, \{ p_{i} (0) \} ) ) ...
\right) /Z $.  
It also satisfies the commutation relation
\begin{equation}
[E(t),E(s)]=2i\hbar f\frac{\partial}{\partial t}\delta(t-s)
\end{equation} 
and has the symmetrized correlation
\begin{equation}
\frac{1}{2}\langle E(t)E(t+\tau)+E(t+\tau)E(t)\rangle = 
\frac{\hbar f}{\pi}\int_{0}^{\infty}\omega
\coth\left[\frac{\beta\hbar\omega}{2}\right]\cos(\omega \tau)d\omega.
\end{equation}
This dynamics yields a classical Langevin equation with Gaussian white noise 
in the classical limit $\hbar\rightarrow 0$.

This dynamics has been applied to the quantum harmonic chain and
investigated some quantum effects in energy transport phenomena 
\cite{ZT90,VR75}. However strictly speaking this quantum Langevin equation 
is the dynamics for a simple particle system in an equilibrium state. 
Therefore this dynamics is not consistent with a nonequilibrium dynamics for
many body system in principle.
The master equation for reduced density matrix (\ref{CTTR})
is derived for general many body system on the assumption that 
only the reservoir is in equilibrium. 
Thus the master equation (\ref{CTTR}) is more suitable in this context.

\section{Energy Transport in the Quantum Harmonic Chain \protect\newline 
at the Weak Coupling Limit}

In this section, 
we investigate energy transport in the quantum harmonic chain 
in contact with two phonon reservoirs at different temperatures
with various types of spectral density of the thermal reservoir
and examine what is common with and what is different from 
the results in the classical case \cite{RLL67} and also the quantum case
\cite{ZT90} with the Ohmic spectral density.
We first discuss the case of weak coupling limit.

\subsection{System}

Here we take the one-dimensional quantum 
harmonic chain,
\begin{eqnarray}
H &=& \sum_{n=1}^{N} \frac{p_{n}^{2}}{2m} 
+ \sum_{n=0}^{N}\frac{m\omega_{0}^2}{2}(x_{n+1}-x_{n})^{2},
\label{HO}
\end{eqnarray}
as the system. This Hamiltonian should be considered to be the renormalized 
Hamiltonian including the second term in (\ref{Hint}).
As in \cite{RLL67}, we impose the fixed boundary condition,
$x_0=x_{N+1}=0$. 
By Fourier transformation:
\begin{eqnarray}
x_{n} = \sqrt{\frac{2}{N+1}}\sum_{k} u_{k} \sin(kn),
\qquad p_{n} = \sqrt{\frac{2}{N+1}}\sum_{k} v_{k} \sin(kn),
\end{eqnarray}
the Hamiltonian is decoupled into the normal modes as
\begin{equation}
H = \sum_{k}\frac{v_{k}^{2}}{2 m} 
+ \frac{m\omega_{k}^{2}u_{k}^{2}}{2},
\end{equation}
where $\omega_k=2\omega_0\sin(k/2)$.
The wave number $k$ runs through the values 
$  k =\frac{\pi\ell}{N+1} \quad\ (\ell = 1,2,\cdots,N-1,N) $.
It is easily found that operators $u_{k}$ and $v_{k'}$ satisfy the 
commutation relations for canonical variables 
\begin{equation}
\left[ u_{k},v_{k'} \right] = i\hbar \delta_{k, k'} 
\quad{\rm and}\quad
\left[ u_{k},u_{k'} \right] = \left[ v_{k}, v_{k'} \right] = 0, 
\end{equation}
and 

Introducing the creation and annihilation operators $a_k^\dagger$ and $a_k$
in the ordinary manner:
\begin{eqnarray}
a_{k} = \sqrt{\frac{m \omega_{0} \sin(k/2) }{\hbar }}
\left( u_{k} + \frac{iv_{k}}{2m\omega_{0} \sin(k/2)} \right)
\quad
{\rm and}
\quad 
a_{k}^{\dagger} = \sqrt{\frac{m \omega_{0} \sin (k/2)}{\hbar }}
\left( u_{k} - \frac{iv_{k}}{2m\omega_{0} \sin(k/2)} \right) ,
\nonumber 
\end{eqnarray}
we obtain
\begin{equation}
 H = \sum_{k} \hbar\omega_{k}
\left( a_{k}^{\dagger} a_{k} + \frac{1}{2}\right).
\label{system}
\end{equation}

\subsection{Equation of Motion of the System which Contacts with 
Two Different Reservoirs}
In order to describe the system whose ends are attached to 
phonon reservoirs at different temperatures, we set dynamical model
where the contacts with thermal baths are taking into account by the 
dissipation terms of the forms in Eq.(\ref{CTTR}). That is, 
variables at the left-end and right-end points $x_1$ and $x_N$ are
linearly coupled with one phonon reservoir at inverse temperature 
$\beta_{\rm L}$ and $\beta_{\rm R}$, respectively.

We assume that the coupling strength $\lambda$ and 
the form of the coupling $\gamma_\alpha$ in Eq.\ (\ref{Hint}) are 
common for both the reservoirs.  
Then the master equation for the reduced density matrix is
written as 
\begin{equation}
\frac{\partial}{\partial t} \rho (t) = 
\frac{1}{i\hbar }\left[ H, \rho(t)\right] 
- \mu\Gamma_{\rm L}\rho(t) 
- \mu\Gamma_{\rm R}\rho(t),
\label{ME2}
\end{equation}
where $\mu=\lambda^2$. In principle the form of the dissipation terms of 
Eq. (\ref{CTTR}) is derived in the condition where the system is coupled only
to one reservoir. Even when the two different reservoirs are contact with 
the system, the decoupled form of the dissipation term in (\ref{ME2}) 
is valid in the order of $\lambda^2$.  
From (\ref{CTTR}) the damping terms $\Gamma_{\rm L}\rho(t)$ and 
$\Gamma_{\rm R}\rho(t)$ are 
\begin{equation}
\Gamma_{\rm L}  \rho (t) = \frac{\pi}{\hbar}
\left\{ \left[ x_{1}, R_{\rm L} \rho (t) \right] 
+ \left[ x_{1}, R_{\rm L}\rho (t) \right]^{\dagger}
\right\}  \quad 
{\rm and }
\quad 
\Gamma_{\rm R} \rho (t) = \frac{\pi}{\hbar}
\left\{ \left[ x_{N}, R_{\rm R}\rho(t) \right] 
+ \left[ x_{N}, R_{\rm R}\rho(t) \right]^{\dagger}
\right\} , \label{TCE}
\end{equation}
respectively. Here operators $R_{\rm L}$ and $R_{\rm R}$ are defined through the
matrix elements
\begin{mathletters}
\begin{eqnarray}
\langle l | R_{\rm L} | m \rangle &=&  
\frac{I_{\rm L} ( \frac{E_{l}-E_{m}}{\hbar} ) 
    - I_{\rm L} (-\frac{E_{l}-E_{m}}{\hbar} )}
{e^{\beta_{\rm L} (E_{l}-E_{m} ) } -1 } 
\langle l | x_{1} | m \rangle ,  \\
\langle l | R_{\rm R} | m\rangle &=&  
\frac{I_{\rm R} ( \frac{E_{l}-E_{m}}{\hbar} ) 
    - I_{\rm R} (-\frac{E_{l}-E_{m}}{\hbar} )}
{e^{\beta_{\rm R} (E_{l}-E_{m} ) } -1 } 
\langle l | x_{N} | m \rangle ,
\end{eqnarray}
\end{mathletters}
and $E_{l}$ denotes the energy eigenvalue for 
state $|l\rangle$. $I_{\rm L}$ and $I_{\rm R}$ are the spectral density
of the left and the right reservoir, respectively.

To solve this equation, we need to express the operators
$R_{\rm L}$ and $R_{\rm R}$ in terms of $a_k$ and $a_k^\dagger$.
The operators $x_{1}$ and $x_{N}$ are 
written as
\begin{equation}
x_{1} = \sqrt{ \frac{\hbar}{2(N+1)m\omega_{0}} }
\sum_{k} \frac{\sin  k }{\sqrt{\sin (k/2)}} 
\left(a_{k}+a_{k}^{\dagger}\right)
\end{equation}
and
\begin{equation}
x_{N}= \sqrt{ \frac{\hbar}{2(N+1)m\omega_{0}} }
\sum_{k} \frac{\sin\left(Nk\right)}{\sqrt{\sin(k/2)}}
\left(a_{k}+a_{k}^{\dagger}\right) .
\end{equation}
Let $|n_{k}\rangle$ denote the eigenstate for the number operator
$a_{k}^{\dagger}a_{k}$ with the eigenvalue $n_{k}$, namely $
 a_{k}^{\dagger}a_{k}|n_{k}\rangle=n_{k}|n_{k}\rangle $.
The eigenstates for the system Hamiltonian (\ref{system}) are
given by the direct product of number states 
$|n_{k}\rangle$ as $|\{n_{k}\}\rangle = \prod_{k} |n_{k} \rangle $,
whose energy eigenvalue is 
$E(\{n_{k}\})=\sum_{k}\left(n_{k}+\frac{1}{2}\right)\hbar\omega_k $.

The matrix elements of $R_{\rm L}$ are given in terms of the eigenstates $|\{n_{k}\}\rangle$ as,
\begin{eqnarray}
\lefteqn{\langle\{n_{k'}\}|R_{\rm L}|\{m_{k'}\}\rangle}\nonumber \\
&=& \sqrt{\frac{\hbar}{2(N+1)m\omega_{0}}}
\sum_{k}  
\frac{\sin k}{\sqrt{\sin (k/2)}} 
\frac{
I_{\rm L}\left(\frac{E(\{n_{k'}\})-E(\{m_{k'}\})}{\hbar}\right)
-I_{\rm L} \left(-\frac{E(\{n_{k'}\})-E(\{m_{k'}\})}{\hbar}\right) 
}
{\exp\{ \beta_{\rm L} 
[E(\{n_{k'}\})-E(\{m_{k'}\})]\}-1 }  
\nonumber \\
& &\qquad\qquad\qquad\quad\times  
\left[ \langle\{n_{k'}\}|a_{k}|\{ m_{k'}\}\rangle 
+ \langle\{n_{k'}\}|a_{k}^{\dagger}|\{ m_{k'}\}\rangle 
\right]. 
\label{RLME}
\end{eqnarray}
Now we note that 
\begin{eqnarray}
\langle\{n_{k'}\}|a_{k}|\{m_{k'}\}\rangle 
&\neq & 0 \quad {\rm only \,\, if} \quad 
n_{k'} = m_{k'} - \delta_{k',k} \quad{\rm for}\quad \forall k', \\
\langle\{n_{k'}\}|a_{k}^{\dagger}|\{ m_{k'}\}\rangle 
&\neq & 0 \quad {\rm only \,\, if}\quad
n_{k'} = m_{k'} + \delta_{k',k} \quad{\rm for}\quad \forall k', 
\end{eqnarray}
and $I(\omega)=0$ for $\omega<0$.  
Then Eq.\ (\ref{RLME}) is transformed into
\begin{eqnarray}
\langle\{n_{k'}\}|R_{\rm L}|\{m_{k'}\}\rangle
&=&\sqrt{\frac{\hbar}{2(N+1)m\omega_{0}}}
\sum_{k}  \frac{I_{\rm L} (\omega_k)\sin k}{\sqrt{\sin(k/2)}}\nonumber \\
& &\mbox{}\times\left[
\frac{e^{\beta_{\rm L}\hbar\omega_{k}}}
{e^{\beta_{\rm L}\hbar\omega_{k}}-1} 
 \langle\{n_{k'}\}|a_{k}|\{ m_{k'}\}\rangle 
+ \frac{1}{e^{\beta_{\rm L}\hbar\omega_{k}}-1} 
 \langle\{n_{k'}\}|a_{k}^{\dagger}|\{m_{k'}\}\rangle\right].
\end{eqnarray}
Thus the operator $R_{\rm L}$ can be represented as
\begin{eqnarray}
R_{\rm L} &=& \sqrt{\frac{\hbar}{8(N+1)m\omega_{0}}}
\sum_{k}\frac{\sin k}{\sqrt{\sin(k/2)}}  
\frac{I_{\rm L} (\omega_{k})}
{\sinh\left(\beta_{\rm L}\hbar\omega_{k}/2\right) } 
\left( e^{ \beta_{\rm L}\hbar\omega_{k}/2}a_{k}
      + e^{-\beta_{\rm L}\hbar\omega_{k}/2}a_{k}^{\dagger}\right) .
\label{Rl}
\end{eqnarray}
In the same manner, the operator $R_{\rm R}$ is represented as
\begin{eqnarray}
R_{\rm R} &=& \sqrt{\frac{\hbar}{8(N+1)m\omega_{0}}}
\sum_{k} \frac{\sin(Nk)}{\sqrt{\sin(k/2)}}  
\frac{I_{\rm R} (\omega_{k} )  }
{\sinh \left( \beta_{\rm R}\hbar\omega_{k} /2 \right) } 
\left( e^{ \beta_{\rm R}\hbar\omega_{k}/2}a_{k}
      + e^{-\beta_{\rm R}\hbar\omega_{k}/2}a_{k}^{\dagger}  \right) .
\end{eqnarray}

\subsection{Moments in the Stationary State}
As will be shown later, to evaluate mean kinetic energy of a particle 
and energy flux in the stationary state, 
we have only to calculate the second moments
\begin{equation}
\label{S1T} 
\langle a_{k} a_{k'}\rangle={\rm Tr}(a_{k}a_{k'}\rho_{\rm st})
\end{equation}
and
\begin{equation}
\label{S2T}
\langle a_{k}^{\dagger} a_{k'}\rangle= 
{\rm Tr} (a_{k}^{\dagger}a_{k'} \rho_{\rm st}),
\end{equation}
where $\rho_{\rm st}$ denotes the stationary solution of Eq.\ (\ref{ME2}).
First, we consider Eq.\  (\ref{S1T}). Because 
the left hand side of Eq.\ (\ref{ME2}) vanishes in the stationary state,
we obtain 
\begin{eqnarray}
\frac{1}{i\hbar } {\rm Tr} 
\left(a_{k}a_{k'} \left[H,\rho\right]\right) 
-\frac{\pi\mu}{\hbar}
\left[
{\rm Tr} \left(a_{k}a_{k'} 
 \left[ x_{1}, R_{\rm L}\rho_{\rm st} \right] \right)
+{\rm Tr} \left( a_{k} a_{k'} 
  \left[ x_{1}, R_{\rm L}\rho_{\rm st}  \right]^{\dagger}  \right)
\right] & & \nonumber \\
\mbox{}-\frac{\pi\mu}{\hbar}
\left[ 
{\rm Tr} \left( a_{k} a_{k'} 
 \left[ x_{N}, R_{\rm R}\rho \right] \right)
+{\rm Tr} \left( 
a_{k} a_{k'}\left[ x_{N}, R_{\rm R}\rho \right]^{\dagger}  
\right)\right]  &=& 0 .
\end{eqnarray}
This equation is rewritten as follows 
after tedious but straightforward calculations,
\begin{eqnarray}
\lefteqn{i(\omega_{k}+\omega_{k'}) \langle a_{k}a_{k'}\rangle  
+ \frac{\pi\mu}{4(N+1)m \omega_{0}}
\sum_{k_{1}}
\frac{\sin k_{1}}{\sqrt{\sin ( k_{1}/2) }} 
\frac{I_{\rm L}(\omega_{k_{1}})}{\sinh(\beta_{\rm L}\hbar\omega_{k_{1}}/2)}} 
\nonumber \\
& &\times\left\{ 
\frac{\sin k'}{\sqrt{ \sin ( k'/2 ) }} 
\left[ e^{\beta_{\rm L}\hbar\omega_{k_{1}}/2} 
(\langle a_{k}a_{k_{1}}\rangle -\langle a_{k_{1}}^{\dagger}a_{k} \rangle)+
e^{-\beta_{\rm L}\hbar\omega_{k_{1}}/2} 
(\langle a_{k} a_{k_{1}}^{\dagger} \rangle - \langle a_{k_{1}}a_{k} \rangle)
\right]\right. \nonumber \\
& & + \left. 
\frac{\sin k}{\sqrt{ \sin (k/2) }} 
\left[ e^{\beta_{\rm L}\hbar\omega_{k_{1}}/2} 
(\langle a_{k'}a_{k_{1}}\rangle - \langle a_{k_{1}}^{\dagger} a_{k'}\rangle) 
+e^{-\beta_{\rm L}\hbar\omega_{k_{1}}/2} 
(\langle a_{k'}a_{k_{1}}^{\dagger}\rangle - \langle a_{k_{1}}a_{k'} \rangle) 
\right]\right\}
\nonumber \\
&+& \frac{\pi\mu}{4(N+1)m\omega_{0}}
\sum_{k_{1}}\frac{\sin ( N k_{1})}{\sqrt{\sin(k_{1}/2)}} 
\frac{I_{\rm R} (\omega_{k_{1}})}{\sinh(\beta_{\rm R}\hbar\omega_{k_{1}}/2)}  
\nonumber \\ 
& & \times  
\left\{ 
\frac{\sin(Nk')}{\sqrt{\sin(k'/2)}} 
\left[ e^{\beta_{\rm R}\hbar\omega_{k_{1}}/2} 
(\langle a_{k}a_{k_{1}}\rangle - \langle a_{k_{1}}^{\dagger}a_{k} \rangle)
+e^{-\beta_{\rm R}\hbar\omega_{k_{1}}/2} 
(\langle a_{k} a_{k_{1}}^{\dagger} \rangle - \langle a_{k_{1}}a_{k}\rangle) 
\right]
\right. \nonumber \\
& & + \left. 
\frac{\sin(Nk)}{\sqrt{\sin(k/2)}} 
\left[e^{\beta_{\rm R}\hbar\omega_{k_{1}}/2} 
(\langle a_{k'}a_{k_{1}}\rangle - \langle a_{k_{1}}^{\dagger}a_{k'}\rangle) 
+e^{-\beta_{\rm R}\hbar\omega_{k_{1}}/2} 
(\langle a_{k'}a_{k_{1}}^{\dagger}\rangle - \langle a_{k_{1}}a_{k'} \rangle) 
\right]
\right\}\nonumber \\
&=& 0 \label{EQ1} \, .
\end{eqnarray}
In the same way, Eq.\ (\ref{S2T}) is also transformed into
\begin{eqnarray}
\lefteqn{i\left( \omega_{k'} - \omega_{k} \right) 
  \langle  a_{k}^{\dagger} a_{k'} \rangle 
+\frac{\pi\mu}{4(N+1)m\omega_{0}}
\sum_{k_{1}} 
\frac{\sin k_{1}}{\sqrt{\sin ( k_{1} /2) }} 
\frac{I_{\rm L} (\omega_{k_{1}})}{\sinh(\beta_{\rm L}\hbar\omega_{k_{1}}/2)}} 
\nonumber \\
& &\times \left\{ 
\frac{\sin k}{\sqrt{ \sin (k/2) }} 
\left[e^{\beta_{\rm L}\hbar\omega_{k_{1}}/2} 
(\langle a_{k_{1}}^{\dagger} a_{k'}\rangle-\langle a_{k'}a_{k_{1}}\rangle)
+e^{-\beta_{\rm L}\hbar\omega_{k_{1}}/2} 
(\langle a_{k_{1}}a_{k'}\rangle-\langle a_{k'}a_{k_{1}}^{\dagger}\rangle) 
\right]
\right. \nonumber \\
& &
+ \left. 
\frac{\sin k'}{\sqrt{\sin(k'/2)}} 
\left[  e^{\beta_{\rm L}\hbar\omega_{k_{1}}/2} 
( \langle a_{k}^{\dagger} a_{k_{1}}         \rangle   
 -\langle a_{k_{1}}^{\dagger} a_{k}^{\dagger } \rangle) 
+e^{-\beta_{\rm L}\hbar\omega_{k_{1}}/2} 
(\langle a_{k}^{\dagger}a_{k_{1}}^{\dagger} \rangle   
 -\langle a_{k_{1}} a_{k}^{\dagger} \rangle) 
\right]\right\}
\nonumber \\
&+& \frac{\pi\mu}{4(N+1)m \omega_{0}}
\sum_{k_{1}} 
\frac{\sin(Nk_{1})}{\sqrt{\sin(k_{1}/2)}} 
\frac{I_{\rm R} (\omega_{k_{1}})}{\sinh(\beta_{\rm R}\hbar\omega_{k_{1}}/2)} 
\nonumber \\
& &\times\left\{ 
\frac{\sin(Nk)}{\sqrt{\sin(k/2)}} 
\left[ e^{\beta_{\rm R}\hbar\omega_{k_{1}}/2} 
(\langle a_{k_{1}}^{\dagger} a_{k'} \rangle 
 -\langle  a_{k'} a_{k_{1}} \rangle)
+e^{-\beta_{\rm R}\hbar\omega_{k_{1}}/2} 
(\langle a_{k_{1}}a_{k'} \rangle   
 -\langle  a_{k'}a_{k_{1}}^{\dagger}\rangle) 
\right]\right. \nonumber \\
& & + \left. 
\frac{\sin (N k' )}{\sqrt{ \sin ( k' /2 ) }} 
\left[ e^{\beta_{\rm R}\hbar\omega_{k_{1}}/2} 
( \langle  a_{k}^{\dagger} a_{k_{1}}        \rangle   
 -\langle  a_{k_{1}}^{\dagger} a_{k}^{\dagger} \rangle) 
+e^{-\beta_{\rm R}\hbar\omega_{k_{1}}/2} 
( \langle a_{k}^{\dagger} a_{k_{1}}^{\dagger} \rangle 
 -\langle a_{k_{1}}a_{k}^{\dagger} \rangle)\right]
\right\} \nonumber \\
&=& 0 \, . \label{EQ2}
\end{eqnarray}

\subsection{Total Energy $E_{\rm st}$}

We will solve these equations by perturbation.
Expanding $\langle a_{k}a_{k'}\rangle$ and 
$\langle a_{k}^{\dagger} a_{k'}\rangle$ with respect to $\mu$,
\begin{eqnarray}
\langle a_{k} a_{k'}\rangle &=& 
           \langle a_{k} a_{k'}\rangle_{0}
+\mu       \langle a_{k} a_{k'}\rangle_{1}
+\mu^{2}   \langle a_{k} a_{k'}\rangle_{2}
+ \cdots , \label{O1}\\
\langle a_{k}^{\dagger} a_{k'}^{\dagger} \rangle &=& 
           \langle a_{k}^{\dagger} a_{k'}^{\dagger} \rangle_{0}
+\mu       \langle a_{k}^{\dagger} a_{k'}^{\dagger} \rangle_{1}
+\mu^{2}   \langle a_{k}^{\dagger} a_{k'}^{\dagger} \rangle_{2}
+ \cdots  ,\\
\langle a_{k}^{\dagger } a_{k'}\rangle &=& 
           \langle a_{k}^{\dagger} a_{k'}\rangle_{0}
+\mu       \langle a_{k}^{\dagger} a_{k'}\rangle_{1}
+\mu^{2}   \langle a_{k}^{\dagger} a_{k'}\rangle_{2}
+ \cdots , \\
\langle a_{k} a_{k'}^{\dagger}\rangle &=& 
           \langle a_{k} a_{k'}^{\dagger}  \rangle_{0}
+\mu       \langle a_{k} a_{k'}^{\dagger}  \rangle_{1}
+\mu^{2}   \langle a_{k} a_{k'}^{\dagger}  \rangle_{2}
+ \cdots \label{O4}.
\end{eqnarray}
we consider the relation of each order of $\mu$.
Using the commutation relations,
\begin{eqnarray}
\langle a_{k} a_{k'} \rangle_{n} 
&=&\langle a_{k'} a_{k} \rangle_{n}, \nonumber \\
\langle a_{k}^{\dagger} a_{k'}^{\dagger} \rangle_{n} 
&=&\langle a_{k'}^{\dagger} a_{k}^{\dagger} \rangle_{n}, \nonumber  \\
\langle a_{k} a_{k'}^{\dagger} \rangle_{n} 
&=&\langle a_{k'}^{\dagger} a_{k} \rangle_{n} 
+\delta_{n,0}\delta_{k,k'}.
\label{RL1}
\end{eqnarray}
we obtain the following relations at the zeroth order,
\begin{equation}
(\omega_{k}+\omega_{k'})\langle a_{k}a_{k'}\rangle_{0} = 0 \quad
{\rm and}
\quad 
(\omega_{k'}-\omega_{k})\langle a_{k}^{\dagger}a_{k'}\rangle_{0} = 0.
\end{equation}
Accordingly we have for all $k$ and $k'$
\begin{equation}
 \langle a_{k}a_{k'}\rangle_{0} = 0,
\label{zeroth1}
\end{equation}
and for $k\neq k'$
\begin{equation}
 \langle a_{k}^{\dagger} a_{k'}  \rangle_{0} = 0. 
\label{zeroth2}
\end{equation}
Putting $k=k'$ in the first order equation of $\mu$, we have 
\begin{eqnarray}
\frac{\sin^{2}k \, I_{\rm L} (\omega_{k} ) }
{\sinh(\beta_{\rm L}\hbar\omega_{k}/2)}
\left(
e^{\beta_{\rm L}\hbar\omega_{k}/2}\langle a_{k}^{\dagger} a_{k}\rangle_{0} 
-e^{-\beta_{\rm L}\hbar\omega_{k}/2} \langle a_{k}a_{k}^{\dagger}\rangle_{0} 
\right) 
& &
\nonumber \\
\mbox{}+\frac{\sin^{2}(Nk) \, I_{\rm R} (\omega_{k} ) }
{\sinh(\beta_{\rm R}\hbar\omega_{k}/2)}
\left(
 e^{\beta_{\rm R}\hbar\omega_{k} /2} \langle  a_{k}^{\dagger} a_{k}\rangle_{0} 
-e^{-\beta_{\rm R}\hbar\omega_{k} /2} \langle  a_{k} a_{k}^{\dagger}\rangle_{0}  
\right)
&=& 0 .
\label{FSE}
\end{eqnarray}
Since $\sin^{2}k=\sin^{2}(Nk)\neq 0$, Eq.\ (\ref{FSE}) leads to
\begin{eqnarray} 
\langle a_{k}^{\dagger}a_{k}\rangle_{0} 
&=& \frac{1}{I_{\rm L}(\omega_{k})+I_{\rm R}(\omega_{k})} 
\left[
 \frac{I_{\rm L}(\omega_{k})}{e^{\beta_{\rm L}\hbar\omega_{k}}-1} 
+\frac{I_{\rm R}(\omega_{k})}{e^{\beta_{\rm R}\hbar\omega_{k}}-1} 
\right],
\end{eqnarray}

Here the energy of the system is,
\begin{equation}
E_{\rm st}={\rm Tr}\left(H\rho_{\rm st}\right)= 
\sum_{k} 
\frac{\hbar\omega_{k}}
{I_{\rm L}(\omega_{k}) + I_{\rm R}(\omega_{k})} 
\left[ 
\frac{I_{\rm L}(\omega_{k})}{e^{\beta_{\rm L}\hbar\omega_{k}}-1} 
+\frac{I_{\rm R} (\omega_{k})}{e^{\beta_{\rm R}\hbar\omega_{k}}-1} 
\right].
\end{equation}
In particular, when
$I_{\rm L} (\omega_{k} ) = I_{\rm R}(\omega_{k})$, we find that 
$E_{\rm st}$ is the arithmetic mean between equilibrium energy at inverse 
temperature $\beta_{\rm L}$ and at $\beta_{\rm R}$ regardless of the types of
the spectral density, i.e.
\begin{equation}
E_{\rm st}=
\frac{1}{2}\left[
\frac{{\rm Tr}\left(He^{-\beta_{\rm L}H}\right)}
{{\rm Tr}\, e^{-\beta_{\rm L}H}} 
+\frac{{\rm Tr}\left(He^{-\beta_{\rm R}H}\right)}
{{\rm Tr}\, e^{-\beta_{\rm R}H}} 
\right].
\label{RL}
\end{equation}

\subsection{Kinetic Energy of a Particle and Energy Flux}

Here we compute mean kinetic energy of each particle and energy flux up to
the first order with respect to $\mu$.
The mean kinetic energy of the $n$th particle, $\varepsilon_{n}$ is 
defined by
\begin{equation}
\varepsilon_{n}
=\left\langle\frac{p_{n}^{2}}{2m}\right\rangle
={\rm Tr}\left(\frac{p_n^2}{2m}\rho_{\rm st}\right)  ,
\label{DLE}
\end{equation}
which is expressed in terms of the creation and annihilation operators as
\begin{eqnarray}
\varepsilon_{n} 
&=& \frac{-\hbar \omega_{0} }{N+1}
\sum_{k, k'} \sqrt{\sin(k/2)\sin(k'/2)}\sin(kn)\sin(k'n) \nonumber \\
&& \times 
\left(
  \langle a_{k} a_{k'}        \rangle 
- \langle a_{k} a_{k'}^{\dagger} \rangle     
- \langle a_{k}^{\dagger} a_{k'} \rangle     
+ \langle a_{k}^{\dagger} a_{k'}^{\dagger} \rangle     
\right) . \label{DLE2}
\end{eqnarray}
Substituting the results obtained in the last subsection into 
the above equation, 
we have
\begin{eqnarray}
\varepsilon_{n}
&=& \frac{\hbar \omega_{0}}{N+1}\sum_{k} \sin (k/2) \sin^{2}\left( k n 
\right)
\left( 2\langle a_{k}^{\dagger} a_{k}   \rangle_{0}  + 1    \right) \nonumber \\
&=& \frac{\hbar \omega_{0}}{N+1}\sum_{k} \sin (k/2) \sin^{2}\left( k n \right)
\left[ 
\frac{2}{ I_{\rm L} (\omega_{k} ) + I_{\rm R} (\omega_{k} )}
\left( 
\frac{ I_{\rm L} (\omega_{k} )  }{e^{\beta_{\rm L}\hbar\omega_{k}}-1} 
+
\frac{I_{\rm R} (\omega_{k} )   }{e^{\beta_{\rm R}\hbar\omega_{k}}-1} 
\right)
+1   
\right] 
. \label{LES}
\end{eqnarray}
In the classical limit, Eq.\ $(\ref{LES})$ becomes
\begin{eqnarray}
2\varepsilon_{n} &=&
\frac{2 T_{\rm L}}{\pi}\int_{0}^{\pi} \, d k \, \sin^{2} (kn) 
\frac{I_{\rm L} (\omega_{k} ) }
{ I_{\rm L} (\omega_{k} ) + I_{\rm R} (\omega_{k} )}
+
\frac{2 T_{\rm R}}{\pi}\int_{0}^{\pi} \, d k \, \sin^{2} (kn) 
\frac{I_{\rm R} (\omega_{k} ) }
{ I_{\rm L} (\omega_{k} ) + I_{\rm R} (\omega_{k} )},
\label{LECLG}
\end{eqnarray}
where $T_{\rm L}=\beta_{\rm L}^{-1}$ and $T_{\rm R}=\beta_{\rm R}^{-1}$ are 
the temperature values of the left reservoir and the right reservoir, 
respectively. In the classical limit, $2\varepsilon_{n}$ can be 
interpreted as the temperature at site $n$.
Especially when both the reservoirs are of the same type, namely, 
$I_{\rm L} (\omega_{k} ) = I_{\rm R} (\omega_{k} )$,
we have 
$2\varepsilon_{n} = \frac{T_{\rm L} + T_{\rm R}}{2} $ regardless of
the types of the spectral density. This means completely flat
temperature profile, which was originally found by RLL 
when the both reservoirs are of the Ohmic type \cite{RLL67}. 
On the other hand, Eq. (\ref{LECLG}) shows that the temperature profile 
in integrable systems can be easily changed by controlling the
combination of the types of spectral density 
of reservoirs, so that in the case of 
$I_{\rm L} (\omega_{k} ) \ne I_{\rm R} (\omega_{k} )$, 
the internal temperature deviates from 
$\frac{T_{\rm L} + T_{\rm R}}{2}$ in the classical limit.

We numerically estimate (\ref{LES}) to investigate the general 
feature of temperature profile for various combinations of
spectral densities. 
We present typical temperature profiles in Fig.$1$.
In Fig.1(a) we take the sub-Ohmic type reservoir $I_{\rm L} = \omega^{0.5} 
$ 
for left side, and the super-Ohmic one 
$I_{\rm R} = \omega^{1.5}$ for the right side. In Fig.1(b), 
the converse case, namely, $I_{\rm L} = \omega^{1.5}$ 
and  $I_{\rm R} = \omega^{0.5} $ is considered.
In both the cases, parameters are set to $m = \hbar = \omega_{0} = 1.0$, and 
$T_{\rm L}= 200.0, T_{\rm R}= 50.0$. 
As is known from these figures, temperature deviates from
the internal temperature value in the same direction in 
the vicinity of both the ends. 
As the result the deviated temperature values become close to 
the temperature of the reservoir whose spectral density has larger power.
We numerically confirmed this dependence for many sets of 
spectral densities $(I_{\rm L}, I_{\rm R})$ and temperatures 
$(T_{\rm L}, T_{\rm R})$ including low temperatures.

Energy flux is defined via the equation of continuity. From the master 
equation (\ref{ME2}), time derivative for the energy of the system satisfies
\begin{eqnarray}
\frac{\partial }{\partial t} {\rm  Tr }\left( H \rho (t) \right)  &=& 
- {\rm Tr} \left(\mu H\Gamma_{\rm L}\rho(t) \right)
- {\rm Tr} \left(\mu H\Gamma_{\rm R}\rho(t) \right) .
\end{eqnarray}
The first term in the right-hand side is regarded as incoming energy flux
from the left reservoir and the second term incoming energy flux from the
right reservoir.  We call the former $J_{\rm L}$ and 
the latter $J_{\rm R}$.  In the stationary state, of course 
$J_L^{\rm st}+J_R^{\rm st}=0$ must hold.
We can calculate $J_{\rm L}^{\rm st}$ as follows,
\begin{eqnarray}
J_{\rm L}^{\rm st}
&=& -\frac{\pi\mu}{\hbar} 
{\rm Tr} \left( H\left[ x_{1}, R_{\rm L}\rho \right] 
            +   H\left[ x_{1}, R_{\rm L}\rho \right]^{\dagger}\right)
\nonumber \\
&=& \frac{\pi\hbar\mu}{(N+1)m}\sum_{k}
\frac{I_{\rm L} (\omega_{k} ) I_{\rm R} (\omega_{k} ) }
{ I_{\rm L} (\omega_{k} ) + I_{\rm R} (\omega_{k} )}
\left( \frac{1}{e^{\beta_{\rm L}\hbar\omega_{k}}-1} 
- \frac{1}{e^{\beta_{\rm R}\hbar\omega_{k}}-1}\right).
\end{eqnarray}
If $N\gg 1$, we can replace the summation by an integral and obtain
\begin{equation}
J_{\rm L}^{\rm st}=\frac{\hbar\mu}{m }
\int_{0}^{\pi}
\frac{I_{\rm L} (\omega_{k} ) I_{\rm R} (\omega_{k} ) }
{ I_{\rm L} (\omega_{k} ) + I_{\rm R} (\omega_{k} )}
\sin^2{k}
\left(
\frac{1}{e^{\beta_{\rm L} \hbar\omega_{k}}-1}-
\frac{1}{e^{\beta_{\rm R} \hbar\omega_{k}}-1} 
\right)dk .
\end{equation}
In the classical limit $(\hbar \rightarrow 0)$, $J_{\rm L}^{\rm st}$ 
goes to
\begin{equation}
J_{\rm L}^{\rm st} = \mu C(T_{\rm L} - T_{\rm R} ),
\end{equation}
where
\begin{equation}
C = \frac{1}{ m \omega_{0} }\int_{0}^{\pi} 
\frac{\sin^{2}k}{\sin(k/2)} 
\frac{ I_{\rm L} \bbox(2\omega_{0}\sin(k/2)\bbox) I_{\rm R} \bbox(2\omega_{0}\sin(k/2)\bbox) }
{I_{\rm L} \bbox(2\omega_{0}\sin(k/2)\bbox) +
I_{\rm R} \bbox(2\omega_{0}\sin(k/2)\bbox)}
dk .
\end{equation}
Thus in the classical limit,
energy flux is proportional to the temperature 
difference and independent of the system size regardless of the 
types and the combinations
of the spectral densities of reservoirs.

\section{Finite Coupling}
The master equation (\ref{CTTR}) is justified only in $O(\mu )$. 
However when we study the model with a finite coupling constant,
the quantitative effect of a finite coupling 
inevitably deviates from those of the original model.
Nevertheless, time evolution of
the reduced density matrix has been regarded as describing a variety of 
relaxation processes, and it successfully explained a variety of 
interesting phenomena in real systems \cite{L73,MEH98}.
This shows that the master equation can well approximate the
dynamics in real dissipative system at least qualitatively.
In some case the time evolution of reduced density matrix with a finite 
coupling reproduces quantitatively correct results even for long time 
\cite{KDH97}.

Thus we investigate here the effect of finite coupling 
which is a small but finite value, 
and discuss the behavior of 
temperature profile qualitatively.
In this section, we 
confine ourselves to the case of the same spectral density at both ends,namely
\begin{eqnarray}
I_{\rm L} = I_{\rm R} = I.
\end{eqnarray}
\subsection{Temperature Profile}
We evaluate contributions from higher-order terms 
and find
deviations from the flat temperature profile near the ends of the chain.
We first calculate the first-order coefficients. From 
the first-order equations in (\ref{EQ1}) and (\ref{EQ2}), 
we have for all $k$ and $k'$
\begin{eqnarray}
\langle a_{k} a_{k'}\rangle_{1} 
&=& \frac{i\pi\left[\sin k\sin k'- \sin(Nk)\sin(N k')\right]}
{4(N+1)m\omega_{0}(\omega_{k}+\omega_{k'}) 
\sqrt{\sin(k/2)\sin(k'/2)}} \nonumber \\
&& \times \left[ I(\omega_{k}) 
\left( n_{\rm L}(\omega_{k})-n_{\rm R}(\omega_{k})\right)
+I(\omega_{k'}) 
\left( n_{\rm L}(\omega_{k'})-n_{\rm R}(\omega_{k'})\right)
\right],
\label{KKD2}
\end{eqnarray}
and for $k\neq k'$
\begin{eqnarray}
\langle a_{k}^{\dagger }a_{k'}\rangle_{1} 
&=& \frac{i\pi 
\left[\sin k\sin k'-\sin(Nk)\sin(Nk')\right]}
{4(N+1)m\omega_{0}(\omega_{k}-\omega_{k'}) 
\sqrt{\sin(k/2)\sin(k'/2)}} \nonumber \\
& & \times \left[ 
I(\omega_{k}) 
\left( n_{\rm L}(\omega_{k})-n_{\rm R}(\omega_{k})\right)
+I(\omega_{k'}) 
\left( n_{\rm L}(\omega_{k'})-n_{\rm R}(\omega_{k'})\right)
\right],
\label{KDKD2}
\end{eqnarray}
where $n_{\rm L}(\omega)$ is the Bose-Einstein distribution 
functions at an inverse temperature $\beta_{\rm L}$
\begin{equation}
n_{\rm L}(\omega)=\frac{1}{e^{\beta_{\rm L}\hbar\omega } -1 },
\end{equation}
and $n_{\rm R}(\omega)$ is that at $\beta_{\rm R}$
\begin{equation}
n_{\rm R}(\omega)=\frac{1}{e^{\beta_{\rm R}\hbar\omega } -1 } . 
\end{equation}
Equations (\ref{EQ1}) and (\ref{EQ2}) imply that the first-order coefficients
must be pure imaginary.  On the other hand, 
$\langle a^{\dagger}_{k}a_{k}\rangle_{1}$ must be real at the same time.  
Thus, we have
\begin{equation}
\langle a^{\dagger}_{k} a_{k } \rangle_{1} = 0.
\end{equation}

From Eqs.\ (\ref{EQ1}) and (\ref{EQ2}), it turns out that
if $n\geq 1$ the $(n+1)$st-order terms are computed
via the following equations from the $n$th order terms;
For all $k$ and $k'$,
\begin{eqnarray}
\langle a_{k} a_{k'} \rangle_{n+1} &=& 
\frac{i\pi}{2(N+1)m\omega_0(\omega_{k}+\omega_{k'})}
\sum_{k_{1}}I(\omega_{k_1})
\nonumber \\
& & \times\left[
\frac{\sin k_1\sin k'+\sin (Nk_1)\sin(Nk')}{\sqrt{\sin(k_1/2)\sin(k'/2)}}
\left( \langle a_{k} a_{k_{1}} \rangle_{n} 
     - \langle a_{k_{1}}^{\dagger} a_{k} \rangle_{n} 
\right)
\right. \nonumber \\
& &  \mbox{}+ \left.
\frac{\sin k_1\sin k+\sin(Nk_1)\sin(Nk)}{\sqrt{\sin(k_1/2)\sin(k/2)}}
\left( \langle a_{k'} a_{k_{1}} \rangle_{n} 
     - \langle a_{k_{1}}^{\dagger} a_{k'} \rangle_{n} 
\right) 
\right].
\end{eqnarray}
If $k\neq k'$,
\begin{eqnarray}
\langle a_{k}^{\dagger} a_{k'} \rangle_{n+1} &=& 
\frac{i\pi}{2(N+1)m\omega_0(\omega_{k'}-\omega_{k})}
\sum_{k_{1}}I(\omega_{k_1})
\nonumber \\
& & \times\left[
\frac{\sin k_1 \sin k'+\sin(Nk_1)\sin(Nk')}{\sqrt{\sin(k_1/2)\sin(k'/2)}}
\left( \langle a_{k}^{\dagger} a_{k_{1}} \rangle_{n} 
     - \langle a_{k}a_{k_{1}} \rangle_{n}^* 
\right)
\right. \nonumber \\
& &  \mbox{}+ \left.
\frac{\sin k_1\sin k+\sin(Nk_1)\sin(Nk)}{\sqrt{\sin(k_1/2)\sin(k/2)}}
\left(\langle a_{k_{1}}^{\dagger}a_{k'} \rangle_{n} 
     - \langle a_{k'} a_{k_{1}} \rangle_{n} 
\right) 
\right],
\end{eqnarray}
and $\langle a_{k}^{\dagger}a_{k} \rangle_{n+1}$ is computed through the other
coefficients of the same order as
\begin{eqnarray}
\langle a_{k}^{\dagger} a_{k} \rangle_{n+1} &=&
\frac{1}{2}
\left( \langle a_{k}a_{k}\rangle_{n+1} 
     + \langle a_{k}a_{k} \rangle_{n+1}^* 
\right)  \nonumber \\
&&\mbox{}-\frac{\sqrt{\sin(k/2)}}{4I(\omega_{k})\sin^{2}k}
\sum_{k'\neq k } 
\frac{\sin k'\sin k+\sin(Nk')\sin(Nk)}{\sqrt{\sin(k'/2)}} 
I(\omega_{k'})  \nonumber \\
& & \times 
\left( \langle a_{k'}^{\dagger}a_{k} \rangle_{n+1}
     + \langle a_{k}^{\dagger}a_{k'} \rangle_{n+1}
     - \langle a_{k}a_{k'} \rangle_{n+1} 
     - \langle a_{k}a_{k'} \rangle_{n+1}^* 
\right).
\end{eqnarray}
Because the above equations contain the spectral density, 
we have to specify its functional form.  
As has given in (\ref{SPEDEF}),
we employ the following form for the spectral density
\begin{equation}
I(\omega) = I_{0}\omega^{\alpha} . \nonumber
\end{equation} 

For $\alpha=1.0$ and $\alpha=1.5$, we have computed mean kinetic energy of 
the $n$th particle $\varepsilon_n$ up
to the 20th order, where quantities (\ref{O1})--(\ref{O4}) seem to converge.
For each $\alpha$, the following sets of temperatures of
for the reservoirs are chosen:
(a) $T_{\rm L}=200.0$ and $T_{\rm R}=50.0$, 
(b) $T_{\rm L}=10.0$ and $T_{\rm R}=0.1$ and 
(c) $T_{\rm L}=0.1$ and $T_{\rm R}=0.02$.
These choices of parameters are the same as used in \cite{ZT90} by ZT.  
Other parameters are commonly set as $m=\omega_0=\hbar=1.0$, $\mu=0.1$, 
and $I_0=1/\pi$.  
In the equilibrium state 
at inverse temperature $\beta$,
$\varepsilon_n$ is given by
\begin{equation}
\varepsilon_n=\phi_n(\beta)=\frac{\hbar\omega_0}{N+1}\sum_{l=1}^N
\sin\frac{\pi l}{2(N+1)}\sin^2\frac{\pi l n}{N+1}
\coth\left[\beta\hbar\omega_0\sin\frac{\pi l}{2(N+1)}\right] , 
\end{equation}
and thus the local temperatures $T_n$ is defined by the above function, i.e.
$T_n=1/\phi_n^{-1}(\varepsilon_n)$.

In fig.$2$ and $3$, $\{ T_{n}\}$ are plotted for $\alpha =1.0$ and $1.5$,
respectively. 
All the figures show that higher-order contributions are small except
near the ends of the chain.  In other words, the bulk behavior is unchanged,
where the temperature profile near the ends of the 
chain exhibits various dependencies on the details of the parameters 
$(T_{\rm L}, T_{\rm R}, \alpha)$.
When the reservoirs are Ohmic, 
the temperature profile near the ends are similar to those obtained by 
ZT with the quantum Langevin approach. 

When the reservoirs are Ohmic and temperature is high (Fig.\ 2(a)), 
temperature drops near the
left end which contacts with the hotter reservoir and rises near the other end
contacting with the colder reservoir. This is the same paradoxical 
behavior as found by RLL and also observed by ZT. 
Such behavior disappear when the reservoirs are super-Ohmic (Fig.\ 3(a)).
The second particles from the ends shows monotonic temperature variation.

Figure 2 (c) and Fig.\  3 (c) exhibit temperature profile 
when the temperatures are low where quantum effects are important.  
These two figures almost coincide. 
In both cases, temperatures near the ends are high 
which should be due to the quantum fluctuations. We may say that differences 
in the spectral
densities does not affect the temperature profile at low temperatures.
In the medium temperature cases, Fig.\ 2 (b) and Fig. 3 (b), mixed behavior
of the classical and quantum features are observed.
(See also the figures in reference \cite{ZT90}). 

For $T_{\rm L} = 200.0$ and $T_{\rm R} = 50.0$, temperature deviations 
of the particle $2$ and the particle $(N-1)$ from the mean internal 
temperature are plotted in Fig.\ 4 for various $\alpha$.
There we find that the peculiarity, i.e. 
inversion of temperature near the ends, is observed in the sub-Ohmic 
and Ohmic cases, while it disappears when 
$\alpha \ge \alpha_c (\simeq 1.04)$.

\subsection{Fokker-Planck Equation in the Classical Limit}

In the previous subsection, the temperature profiles
is found to depend on values of $\alpha$.
In particular, the peculiarity found by RLL \cite{RLL67} 
disappears in the super-Ohmic regime.
Since the differences are seen at high temperatures, 
some characteristics depending to the values $\alpha$ must appear 
in the Fokker-Planck equation 
obtained from the master equation in the classical limit.
Actually, we will find that 
the diffusion term in the Fokker-Planck equation
takes a different form from that derived from the Langevin equation
except the Ohmic case.
In order to study the difference in relaxation at the contacting point,
we investigate the Fokker-Planck equation for a system with
a single reservoir.

When a heat reservoir is attached to the left end of the chain,  
the classical Langevin equations for canonical variables $x_n(t)$ and
$p_n(t)$, $n=1,2,...,N$ are
\begin{eqnarray}
\frac{\partial x_{n}}{\partial t} &=& \left\{ x_{n} , H \right\} \\
\frac{\partial p_{n}}{\partial t} &=& \left\{ p_{n} , H \right\}
-\delta_{n,1}\nu\frac{p_n}{m}+\delta_{n,1}\xi(t) 
\label{LGVN}
\end{eqnarray}
where $\left\{\cdot,\cdot\right\}$ means the Poisson bracket.
The correlation function of the Gaussian white random force $\xi(t)$ is
connected with the damping constant $\nu$ and the temperature at the first
particle $\beta$ via the fluctuation-dissipation theorem as
\begin{equation}
\langle \xi(t)\xi(t')\rangle =\frac{2\nu}{\beta}\delta(t-t').
\end{equation}
As is well known, the Langevin equations are equivalent to 
the Fokker-Planck equation 
\begin{eqnarray}
\frac{\partial P(t) }{\partial t} &=& 
\left\{ H, P(t) \right\} 
+ \nu \frac{\partial }{\partial p_{1}} 
\left( \frac{p_{1}}{m} + \beta^{-1}
\frac{\partial }{\partial p_{1}}  \right) P(t),
\label{RLLCF}
\end{eqnarray}
where $P(t)$ is the distribution function on the phase space.

We now turn to our master equation.
Inserting the representation for operator $R_{\rm L}$ $(\ref{Rl})$ 
into the master equation (\ref{ME}), we have
\begin{eqnarray}
\frac{\partial \rho(t)}{\partial t}&=& 
\frac{1}{i\hbar}
[H,\rho(t)]-\frac{\pi\mu}{\sqrt{8(N+1)m\hbar\omega_{0}}} 
\sum_{k} \frac{\sin  k }{\sqrt{\sin (k/2)}}  
\frac{ I(\omega_{k})  }
{\sinh(\beta\hbar\omega_{k}/2) } \\
& & \times \left\{\left[x_{1}, 
\left(e^{\beta\hbar\omega_{k}/2} a_{k}
+ e^{-\beta\hbar\omega_{k} / 2} a_{k}^{\dagger}
\right)\rho(t)\right]-
\left[ x_{1}, 
\rho (t) \left( e^{ \beta\hbar\omega_{k} / 2} a_{k}^{\dagger }
+ e^{-\beta\hbar\omega_{k}/2} a_{k}\right)\right]\right\},
\nonumber
\end{eqnarray}
where we omitted suffix L.
Expressing the creation and annihilation operators by 
the position and momentum operators,
\begin{eqnarray}
\frac{\partial\rho(t)}{\partial t} &=& 
\frac{1}{i\hbar}[H,\rho(t)] \nonumber \\
& & \mbox{}-\frac{\pi\mu}{(N+1)\hbar}\sum_{k,n} 
I(\omega_{k})\sin k \sin(kn)
\left\{
\coth\left(
\frac{\beta\hbar\omega_{k}}{2}
\right)
[x_{n},[x_{1},\rho(t)]]  \right. \nonumber \\
&&\mbox{}+ \left. \frac{i}{2m\omega_{0}\sin(k/2)}
\left(
p_{n}[x_{1},\rho(t)] +[x_{1},\rho(t)]p_{n}+2[x_{1},p_{n}]\rho(t) 
\right) 
\right\}. 
\label{FPEM2}
\end{eqnarray}
In the classical limit, the density matrix $\rho(t)$ is replaced by the 
distribution function $P(t)$
Therefore,
Eq.\ $(\ref{FPEM2})$ is transformed into 
\begin{equation}
\frac{\partial P(t)}{\partial t}= \{H,P(t)\}  + 
\sum_{n=1}^{N}C_{n} \frac{\partial }{\partial p_{1}}
\left( 
\frac{p_{n}}{m}+\beta^{-1}\frac{\partial}{\partial p_{n}}
\right) P(t),
\label{FPours}
\end{equation}
where
\begin{equation}
C_{n} = \frac{2\mu}{\omega_{0}}\int_{0}^{\pi}
I\bbox(2\omega_{0}\sin(k/2)\bbox)\cos(k/2)\sin(kn) dk ,
\label{FECTR}
\end{equation}
The time-evolution equation for the covariance matrix derived from Eq.\ 
(\ref{FPours}) is also confirmed to agree with the classical limit of 
the corresponding quantum equation. 

When the reservoir is Ohmic, namely $I(\omega)=I_0\omega$,
the coefficients $C_{n}$ are evaluated as
\begin{equation}
C_{n}= \pi\mu I_{0}\delta_{n,1},
\end{equation}
and Eq.\ (\ref{FPours}) agrees with the Fokker-Planck equation 
derived from the Langevin equation (\ref{RLLCF}).  
In this case, the two-point function $(\ref{TPF})$ tends to 
the delta function in the classical limit 
\begin{equation}
\lim_{\hbar\rightarrow 0}\Phi(t)=
\frac{2\pi I_{0}}{\beta} \delta (t).
\end{equation}
Therefore, we find that the correlation function of the noise is white in the Ohmic case, 
which is consistent with the Langevin equation (\ref{LGVN}). 

If the reservoir is sub-Ohmic or super-Ohmic, however, $C_n$ does not  
vanish for $n\geq 2$. Figure $5$ shows $C_{n}$ as a function of $n$ 
for various values of $\alpha$. The sign of $C_{n}$ 
($n\geq 2$) is positive in the sub-Ohmic regime and negative in the
super-Ohmic regime.  The difference in temperature profiles discussed in
the previous subsection should be explained by this $\alpha$-dependence of the
coefficients $C_n$.

\section{Summary and Discussion}

We investigated the effect of the types of reservoirs on 
the thermal conduction in the harmonic chain. 
We derived the master equation for general many body system 
in contact with phonon reservoirs. 
In many body system, the dissipation term is different from
one of one-particle system due to the noncommutability of many body 
interaction, so that the dissipation term has rather complicated form. 
However we have the explicit form for the
dissipation term (\ref{RFORM}) and (\ref{CTTR}).
We used it as the basic equation to study behavior of the system. 
The equation generally satisfies the necessary condition for the 
master equation that the 
canonical distribution must be a stationary solution when the reservoirs
are at the same temperature. 

In Sec.\ 3, we have applied the master equation to energy transport in 
the quantum harmonic chain.  We attached a phonon reservoir at one end 
and another at the other end.  At weak coupling limit 
$(\lambda\rightarrow 0)$, we obtained explicit form of internal energy 
and energy flux. We rigorously proved that
when the spectral densities of the reservoirs are of same type,
the total energy of the system takes the arithmetic mean of 
the equilibrium energies at $T_{\rm L}$ and $T_{\rm R}$ regardless of
the type of the spectral density. This result leads to the classical
temperature $\frac{T_{\rm L} + T_{\rm R}}{2}$ which is originally found 
by RLL using the Ohmic type of reservoir.
On the other hand, when the types of spectral densities
are different, the internal temperature is a function of the both densities, 
so that the temperature does not converge to $\frac{T_{\rm L} + T_{\rm R}}{2}$
in the classical limit. The difference of spectral densities induces the
deviation of temperatures around the both edges from the internal value.
The temperature in the vicinity of both ends become close to the temperature
of the reservoir whose spectral density has larger power. 
We numerically confirmed that this feature is general when the reservoirs 
are of different types.

We numerically investigated the effect of finite coupling.
We considered only the case of the same spectral densities of the reservoirs. 
Finite coupling contributes to the temperature profile 
only near the ends of the chain and bulk
behavior is the same as that of weak coupling limit.
We found that the profile near the ends depends 
on the spectral density for the reservoirs.  
When the reservoir temperature is sufficiently low where quantum
fluctuations are dominant, temperature growth near both the ends was
observed in every case. When the reservoir
temperatures are high enough and the reservoir is sub-Ohmic or Ohmic, 
the same peculiar behavior, i.e., nonmonotonic change of
the temperature, is observed as found in \cite{RLL67}. 
However, in the case of super-Ohmic reservoir, the peculiarity 
disappears.  

In order to understand the dependence on the spectral density,
we derived Fokker-Planck equations from the master equation 
in the classical limit.  If the reservoir is Ohmic the Fokker-Planck 
equation agrees with the 
standard one derived from the Langevin equation.
When the reservoir is non-Ohmic, however, there appears difference in the 
diffusion term , i.e., form of second derivative.
The coefficients of the diffusion terms were calculated from the 
spectral density.This difference causes different 
temperature profiles near the ends of the chain.

We expect that the master equation derived here can be used for other 
systems such as spin systems for which the Langevin approach is 
practically difficult.  
In the case of the harmonic chain, operator $R$ was written in a simple form 
by using some system operators.  Thus we were able to analyze the master 
equation systematically.  This can be done because the harmonic chain is
integrable.  Thus, similar procedure can be developed for other integrable 
systems, e.g.\ the XY model \cite{STMPP}. 

In this paper we have confined ourselves to the integrable system. However the
master equation derived here is generally applicable to any system
because the matrix element of $R$ is explicitly given.
Thus it would be an interesting future problem to study the thermal 
conductivity in nonintegrable 
system where the Fourier heat law is realized \cite{STM96}.  

\section*{Acknowledgment}
We gratefully acknowledge partial financial 
support from Grant-in-Aid for Scientific
Research from the Ministry of Education, Science and Culture.

\begin{figure}
\caption{Temperature profile along the chain for 
$T_{\rm L}=200.0, T_{\rm R}= 50.0$:  
(a) $I_{\rm L}(\omega ) = \omega^{0.5}$ and 
$I_{\rm R}(\omega ) = \omega^{1.5}$.;
(b)  $I_{\rm L}(\omega ) = \omega^{1.5}$ and 
$I_{\rm R}(\omega ) = \omega^{0.5}$. The system size is $N=150$.}
\end{figure}

\begin{figure}
\caption{Temperature profile along the chain for $\alpha = 1.0$:
(a) $T_{\rm L}=200.0, T_{\rm R}= 50.0 $; (b) $T_{\rm L}=10.0, T_{\rm R}= 0.1$;
(c) $T_{\rm L}= 0.1, T_{\rm R}= 0.02 $.  The system size is $N=150$.}
\end{figure}

\begin{figure}
\caption{
Temperature profile along the chain for $\alpha = 1.5$:
(a) $T_{\rm L}=200.0, T_{\rm R}= 50.0 $; (b) $T_{\rm L}=10.0, T_{\rm R}= 0.1$;
(c) $T_{\rm L}= 0.1, T_{\rm R}= 0.02$.  The system size is $N=150$.} 
\end{figure}

\begin{figure}
\caption{
Deviations of the temperature at particle $2$ and particle $(N-1)$
from the mean internal temperature $T_{\rm av}$.  The temperatures
of the reservoirs are $T_{\rm L}=200.0$ and $T_{\rm R} = 50.0$.
Thus, $T_{\rm av} = 125.0$.}
\end{figure}

\begin{figure}
\caption{
Coefficients $C_{n}$ as a function of $n$ for various values of 
$\alpha$.}
\end{figure}

\newpage
\begin{center}
\begin{figure}[htbp]
\noindent
\epsfxsize=10cm \epsfbox{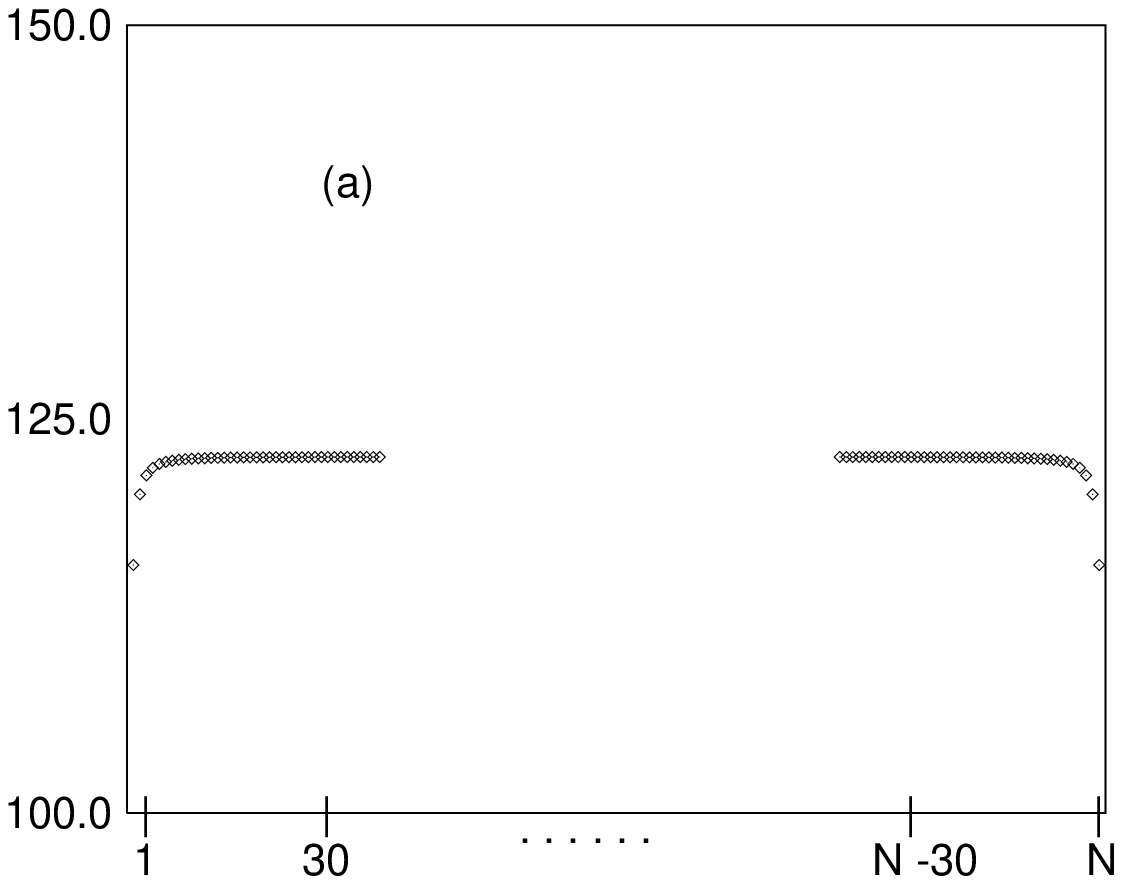}
\vspace*{0.5cm}
\epsfxsize=10cm \epsfbox{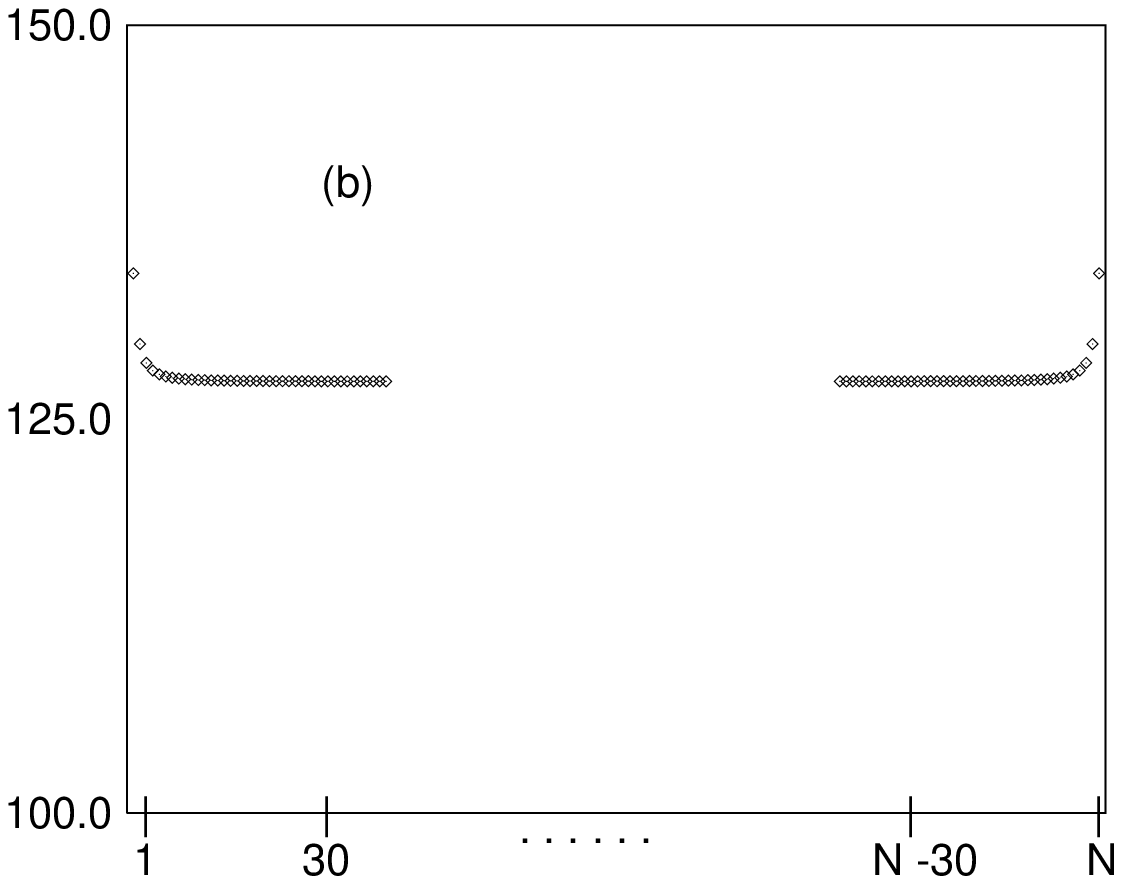}\\
\vspace*{0.5cm}
{\rm \Large Fig.1,  K. Saito, S.Takesue, S. Miyashita}
\end{figure}
\end{center}

\newpage 
\begin{center}
\begin{figure}[htbp]
\noindent
\epsfxsize=10cm \epsfbox{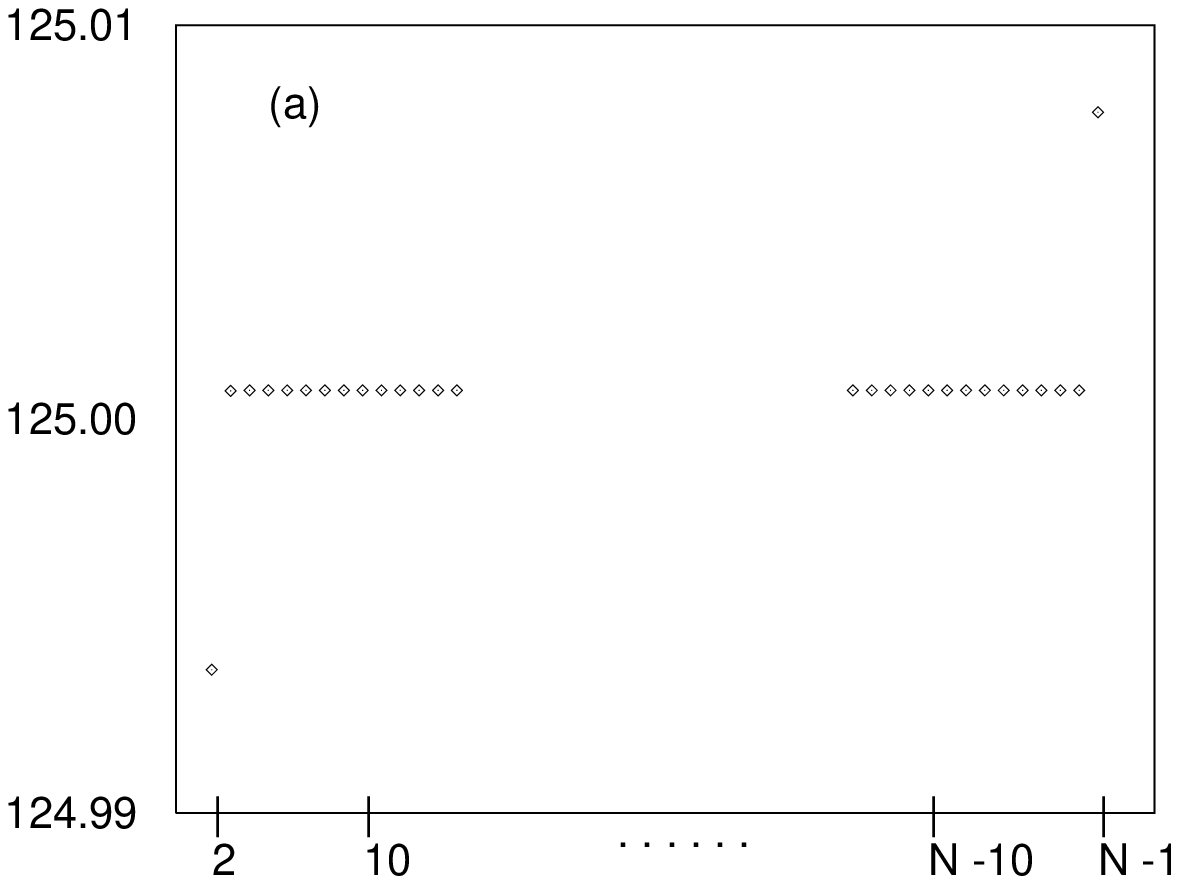} 
\vspace*{0.5cm}
\epsfxsize=10cm \epsfbox{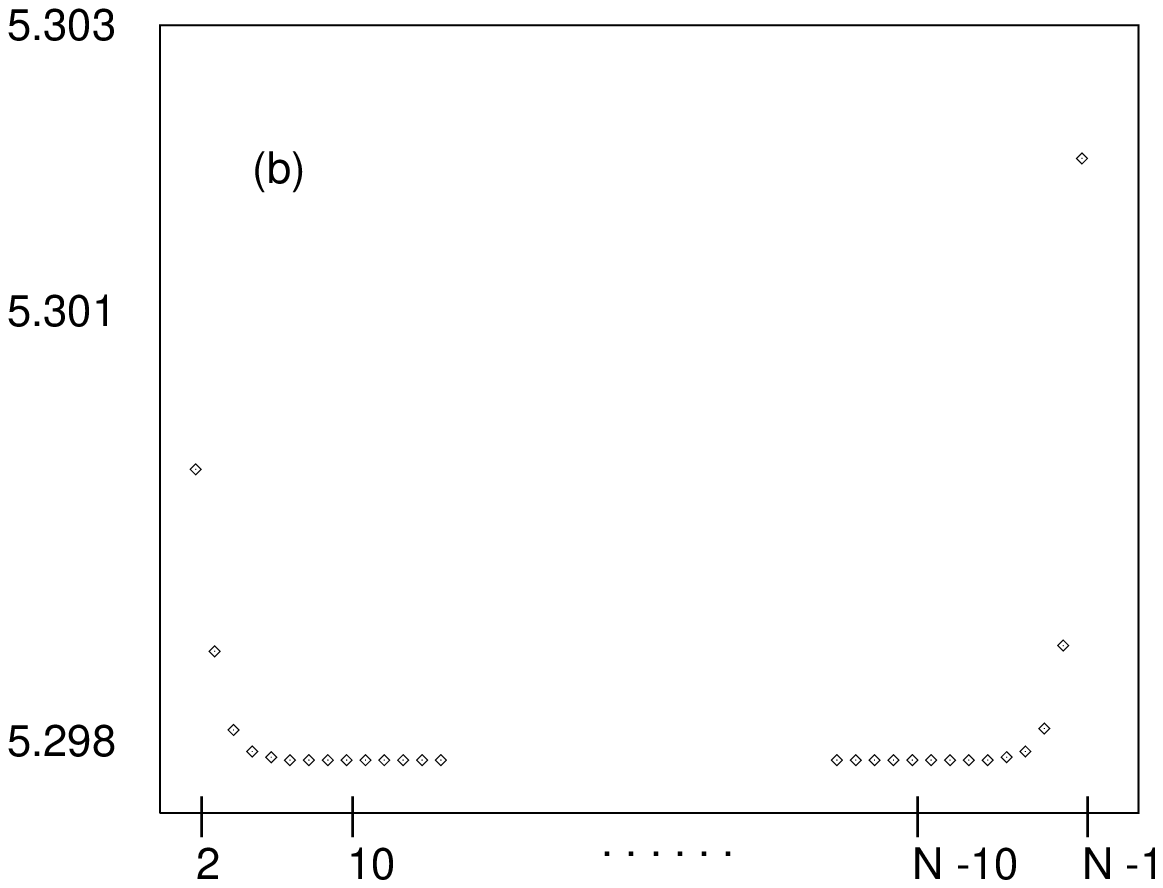} 
\vspace*{0.5cm}
\epsfxsize=10cm \epsfbox{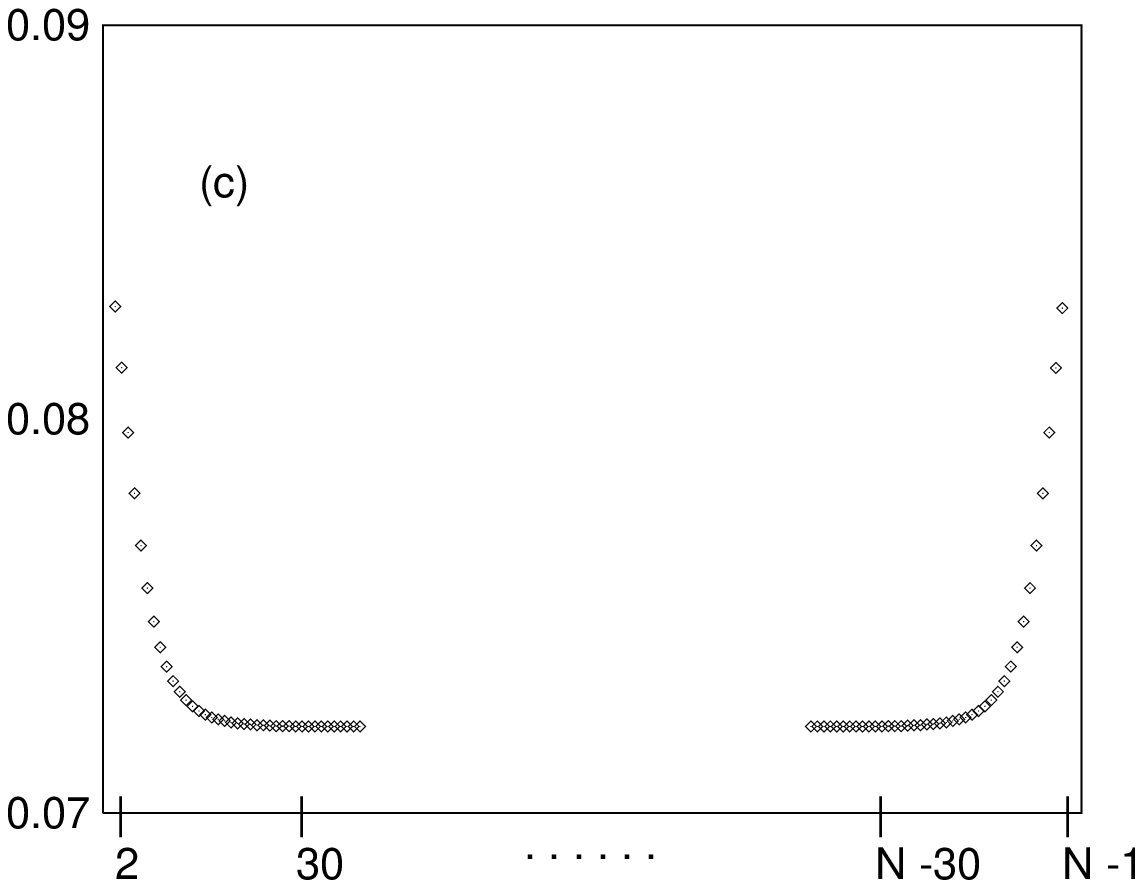}  \\
\vspace*{0.5cm}
{\rm \Large Fig.2,  K. Saito, S.Takesue, S. Miyashita}
\end{figure}
\end{center}
\newpage 
\begin{center}
\begin{figure}[htbp]
\noindent
\epsfxsize=10cm \epsfbox{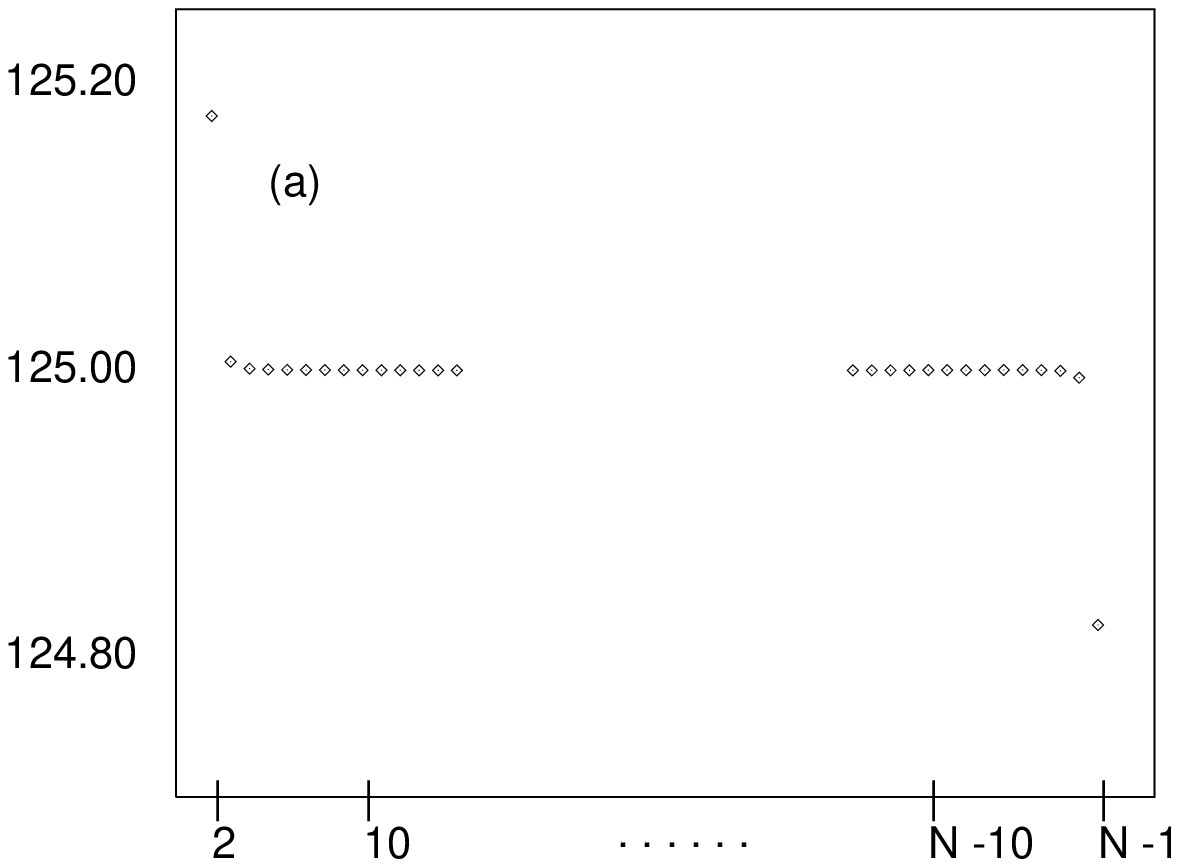}
\vspace*{0.5cm}
\epsfxsize=10cm \epsfbox{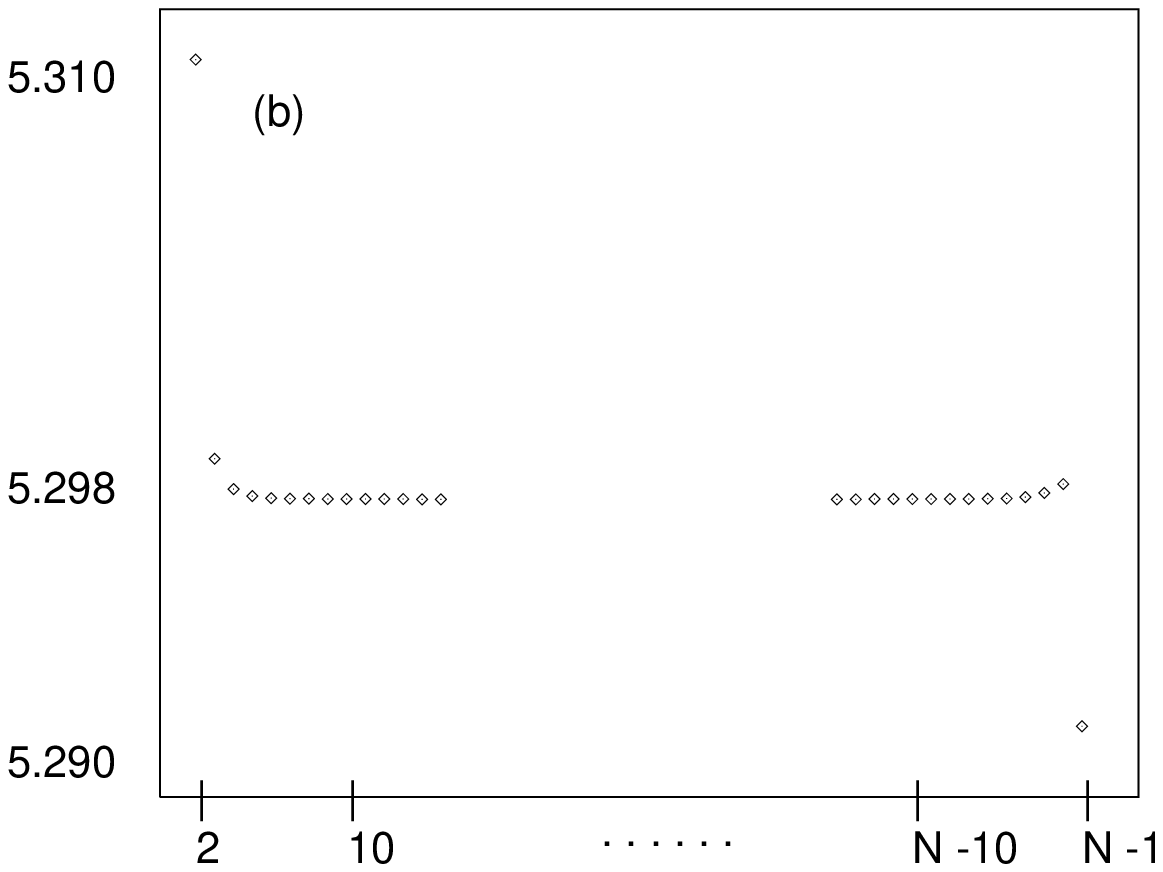}
\vspace*{0.5cm}
\epsfxsize=10cm \epsfbox{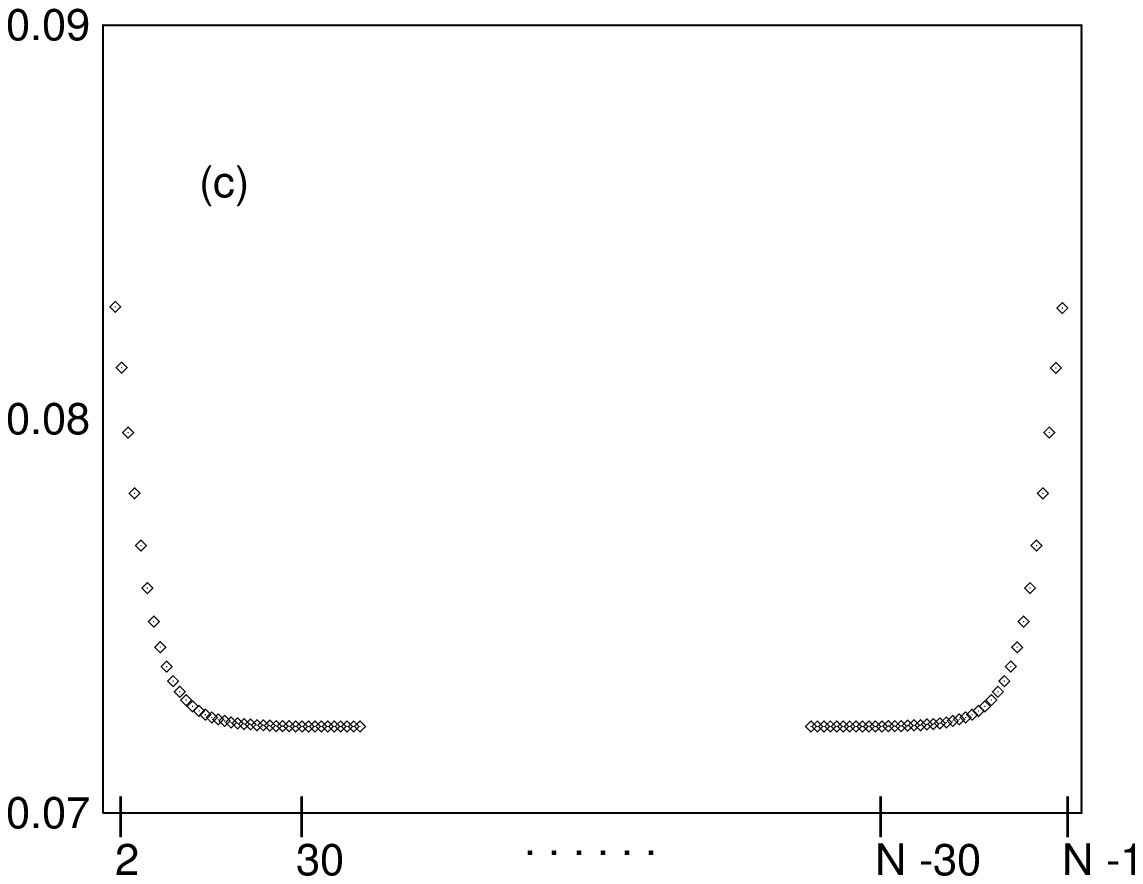} \\
\vspace*{0.5cm}
{\rm \Large Fig.3,  K. Saito, S.Takesue, S. Miyashita}
\end{figure}
\end{center}
\newpage 
\begin{center}
\begin{figure}[htbp]
\noindent
\epsfbox{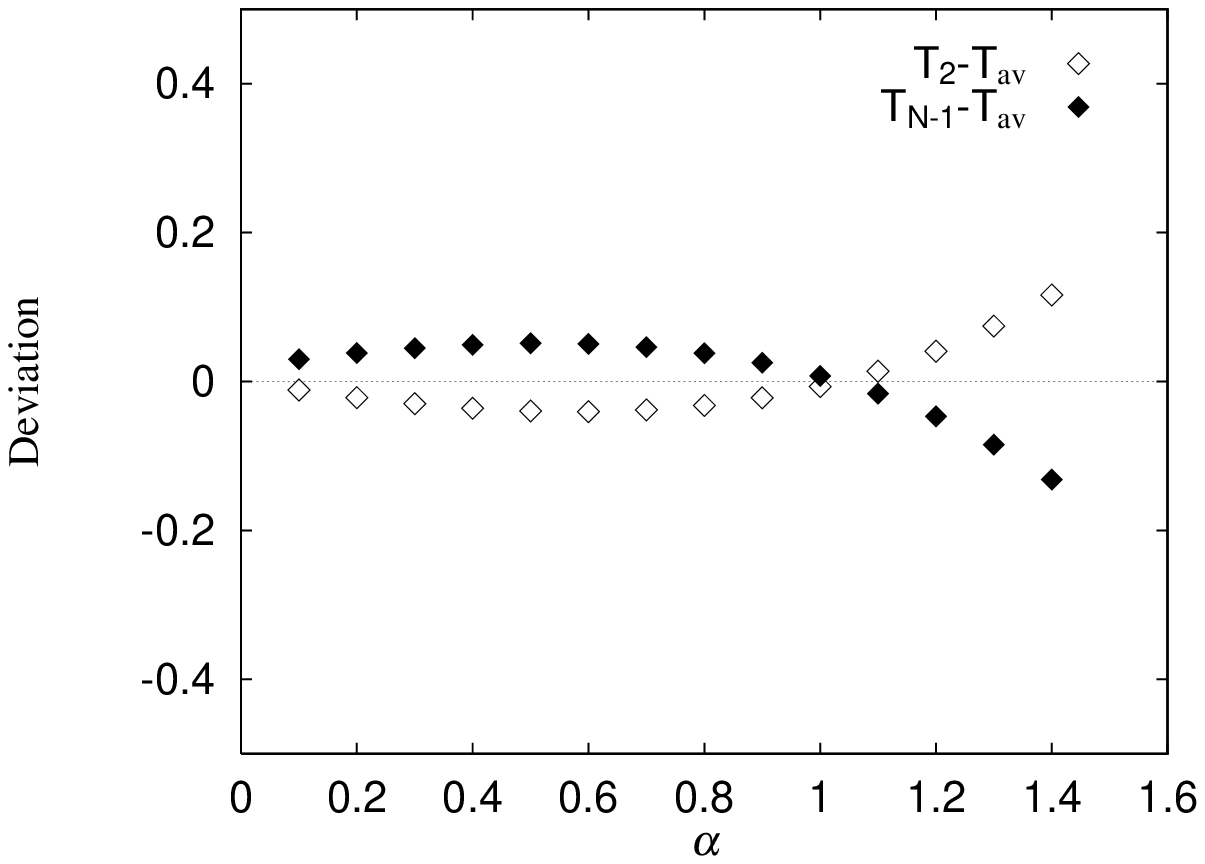} \\
\vspace*{1.5cm}
{\rm \Large Fig.4,  K. Saito, S.Takesue, S. Miyashita}
\end{figure}
\end{center}
\newpage 
\begin{center}
\begin{figure}[htbp]
\noindent
\epsfbox{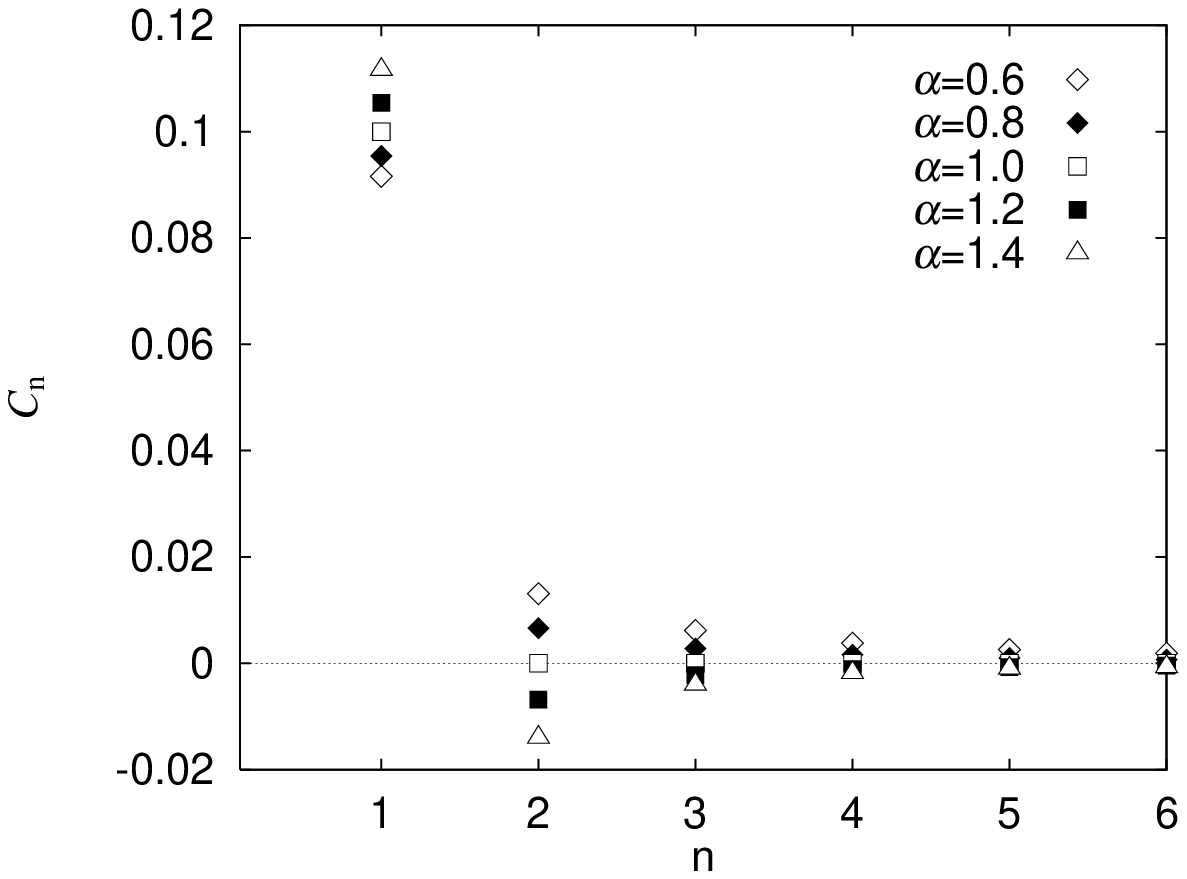} \\
\vspace*{1.5cm}
{\rm \Large Fig.5,  K. Saito, S.Takesue, S. Miyashita}
\end{figure}
\end{center}

\end{document}